\documentclass[aps,prb,twocolumn,superscriptaddress,10pt]{revtex4-2}
\pdfoutput=1 
\usepackage[utf8]{inputenc}
\usepackage[english]{babel}
\usepackage[T1]{fontenc}
\usepackage[pdftex]{graphicx}
\usepackage{amsmath}
\usepackage{amsthm}
\usepackage{amssymb}
\usepackage{braket}
\usepackage{xcolor}
\usepackage{dcolumn}
\usepackage{bbm}
\usepackage{bbold}
\usepackage[normalem]{ulem}
\usepackage[colorlinks=true,linkcolor=blue,citecolor=blue,urlcolor=blue,unicode]{hyperref}
\usepackage{footmisc}

\begin{document}
\title{Weak integrability breaking perturbations of integrable models}
\author{Federica Maria Surace}
\affiliation{Department of Physics and Institute for Quantum Information and Matter,
California Institute of Technology, Pasadena, California 91125, USA}
\author{Olexei Motrunich}
\affiliation{Department of Physics and Institute for Quantum Information and Matter,
California Institute of Technology, Pasadena, California 91125, USA}

\begin{abstract}
A quantum integrable system slightly perturbed away from integrability is typically expected to thermalize on timescales of order $\tau\sim \lambda^{-2}$, where $\lambda$ is the perturbation strength. We here study classes of perturbations that violate this scaling, and exhibit much longer thermalization times $\tau\sim \lambda^{-2\ell}$ where $\ell>1$ is an integer.  Systems with these ``weak integrability breaking'' perturbations have an extensive number of quasi-conserved quantities that commute with the perturbed Hamiltonian up to corrections of order $\lambda^\ell$. We demonstrate a systematic construction to obtain families of such weak perturbations of a generic integrable model for arbitrary $\ell$. We then apply the construction to various models, including the Heisenberg, XXZ, and XYZ chains, the Hubbard model, models of spinless free fermions, and the quantum Ising chain. Our analytical framework explains the previously observed evidence of weak  integrability breaking in the Heisenberg and XXZ chains under certain perturbations.
\end{abstract}

\maketitle

\section{Introduction}
Understanding how a many-body system reaches thermal equilibrium is one of the fundamental questions in statistical mechanics.
A generic (non-integrable) quantum many-body system that evolves under unitary dynamics is typically expected to thermalize \cite{Deutsch1991ETH, Srednicki1994ETH, rigol2008thermalization, DAlessio2016, gogolin2016equilibration, Mori_2018}:
When thermalization occurs, expectation values of local observables reach stationary values that depend only on few properties of the initial state and can be predicted with usual statistical mechanics ensembles.
An exception is represented by integrable models:
These models have a large number of extensive local conserved quantities that retain information about the initial state.
As a consequence, integrable models do not thermalize in the usual sense, but, in contrast, their time evolution can be described as a relaxation to a stationary ensemble that includes all conserved quantities, called {\it generalized Gibbs ensemble} (GGE) \cite{kinoshita2006quantum,Rigol2007,Cassidy2011,Gogolin2011,Fagotti2014,Essler_2016}.

In the presence of a perturbation that breaks integrability, usual thermalization is again expected to take place.
However, if the perturbation is small, the process may require a long time.
On a finite time scale, the dynamics is approximately described by the evolution under the integrable unperturbed Hamiltonian.
The system initially relaxes to a stationary state of the unperturbed Hamiltonian ({\it prethermalization}), while genuine thermalization only occurs at later times \cite{Berges2004,Kollar2011,Langen_2016}.
This later thermalization is typically modelled using Fermi's golden rule, which prescribes a thermalization time $\tau \sim \lambda^{-2}$, where $\lambda$ is the perturbation strength \cite{Mallayya2019,Mallayya2021}.
However, for specific Hamiltonians this timescale can be much longer.
For example, Abanin {\it et.~al.} \cite{abanin2017rigorous} proved that for unperturbed Hamiltonians with special structure producing equally-spaced sectors such as the Hubbard model in the limit of zero hopping, the thermalization time has a lower bound $\tau \sim e^{c/\lambda}$.  
The result can be essentially proven by using local unitary (Schrieffer-Wolff) transformations that, order by order, eliminate the perturbations in the rotated frame, stopping at a ``sweet'' order where the combinatorial growth of the number of perturbation terms becomes overpowering factor.
(The rigorous theory of prethermalization was also proved for unperturbed Hamiltonians with energy sectors determined by more than one frequency~\cite{Else2020,DeRoeck2019}.)
With this approach, the conserved quantities of the unperturbed Hamiltonian that label the equally-spaced sectors (such as the doublon number in the Hubbard model) are ``dressed'' by the perturbation, and are conserved up to times that are exponentially large in $1/\lambda$.
There is no analog of this rigorous theory of prethermalization for general unperturbed Hamiltonians, and it is not possible, in general, to find such unitary Schrieffer-Wolff transformations in the thermodynamic limit.
Nevertheless, the emergence of approximate conserved quantities was observed in certain other models, that do not belong to this special class.

More specifically, Kurlov~{\it et.~al.}~\cite{Kurlov2022} recently showed that the Heisenberg chain perturbed with a next-nearest neighbour SU$(2)$ symmetric interaction has several approximately conserved quantities that commute with the perturbed Hamiltonian up to corrections of order $\lambda^2$.
A similar case was observed for the Heisenberg and XXZ chains perturbed with isotropic next-nearest neighbor interaction by Jung~{\it et.~al.} in an earlier  Ref.~\cite{Jung2006}, where the authors found an anomalously large heat conductivity that they explained from the presence of an approximate conservation law.
In such cases, the presence of approximate conservation laws can lead to longer thermalization times, of the order of $\tau \sim \lambda^{-4}$.
These approximately conserved quantities were discovered through the use of numerical methods or heuristic procedures, and are typically limited to a small number of operators.
It remained unclear why approximate conservation laws appear in these systems.

In this work we use a systematic analytical approach to find models with approximately conserved quantities, and to compute such quantities.
This approach is based on recently studied so-called long-range deformations of integrable spin chains. 
These deformations were introduced in \cite{Bargheer_2008,Bargheer_2009} in the context of AdS/CFT.
It was later shown that some of them can be seen as generalizations of $T\overline T$ deformations of $1+1$-dimensional integrable quantum field theories~\cite{balazs2020tt,Marchetto2020,Doyon2022}.
One can view these deformations as produced by continuous unitary transformations generated by special operators which are not necessarily local but which produce local terms order-by-order in the expansion in the continuous parameter. 
These special operators (that include so-called boosted operators or bi-local operators) are constructed from the original integrals of motion.

In our work, we utilize the idea that truncations of these deformations at finite order can be viewed as special perturbations that break integrability but more weakly than generic perturbations \cite{Pozsgay2020, Szasz2021}; in what follows, we will often refer to such weak integrability breaking perturbations as ``weak perturbations'' in short (an alternative name which we will sometimes use is ``nearly-integrable''~\cite{Durnin2021}).
Since the deformations are generated by unitary flows, the same transformations apply to all integrals of motion of the original integrable model, and the corresponding truncations then produce approximate conserved quantities [which we will often refer to as quasi-conserved quantities or quasi-integrals of motion (quasi-IoMs)] of the perturbed model.
We show that the Heisenberg and XXZ chains perturbed by the second-neighbor Heisenberg term can be cast as an example of this approach.
This explains the previous findings of few quasi-integrals of motion in these chains and also shows that all integrals of motion of the original integrable chains give rise to quasi-integrals of motion of the perturbed chain, allowing us to derive  all of them.

The power of this approach is that it allows construction of large families of weak perturbations to a given integrable model.
Besides the Heisenberg and XXZ models, we show additional simple examples for a number of well-known integrable models including the Hubbard model, free-fermion models, and quantum Ising models.
Some of these examples --- like density-assisted hopping perturbation for spinless fermions or a particular self-dual deformation of the quantum Ising chain --- have in fact appeared in the literature in various contexts without appreciating that they are weak integrability breaking perturbations.

As another application of our systematic approach, we provide an explicit demonstration how to construct weak perturbations beyond leading order: 
Thus, we show that the Heisenberg chain with second-neighbor and third-neighbor interactions (with specific coupling proportional to the square of the second-neighbor coupling) is a second-order weak integrability breaking perturbation.

The existence of approximate conserved quantities implies that the relaxation to thermal equilibrium upon the introduction of a weak perturbation is slow:
We provide a rigorous bound on the thermalization rate that is linear in the effective strength of the remaining generic integrability-breaking perturbation.
We also review and verify a non-rigorous ``Fermi-Golden-Rule'' type estimate which is believed to be quadratic in the remaining perturbation strength.
These types of estimates apply to all weak integrability breaking perturbations considered here.

The paper is organized as follows.
In Section~\ref{sec:deform} we review the families of long-range deformations of integrable models introduced in Refs.~\cite{Bargheer_2008,Bargheer_2009}, and the classes of operators (extensive, boosted, bilocal, or a combination of the three) that generate them. 
These deformations depend smoothly on a parameter $\lambda$.
In Section~\ref{sec:finite} we consider truncations of the deformations to a finite order in $\lambda$, and thus derive families of Hamiltonians with quasi-conserved quantities.
In Section~\ref{sec:ex_boost} we examine several examples of weak perturbations generated by boosted operators to first order in $\lambda$: We apply the construction to the Heisenberg, XYZ, XXZ chains and to the Hubbard model.
In Section~\ref{sec:ex_biloc} we consider additional examples of first-order weak perturbations, that are generated by bilocal operators:
We focus on free spinless fermions, the quantum Ising chain, and the Heisenberg chain.
In Section~\ref{sec:ex_higher} we demonstrate how to apply the procedure to obtain weak perturbations beyond first order, and we consider for concreteness the case of the Heisenberg chain. 
In Section~\ref{sec:therm} we discuss how weak perturbations imply parametrically longer thermalization times.
Finally, in Section~\ref{sec:conclusions} we summarize our conclusions and suggest future outlooks.

\section{Deformation of local conserved charges}
\label{sec:deform}
We consider a one-dimensional quantum system whose Hamiltonian $H$ commutes with a set of extensive local operators ({\it charges}) $\{Q_\alpha\}$ of the form
\begin{equation}
Q_\alpha = \sum_j q_{\alpha,j},
\end{equation}
where $j$ labels the sites of a one-dimensional lattice and the charge density $q_{\alpha,j}$ is an operator with finite range (i.e., acting on a finite number of sites around $j$).
We also assume that $[Q_\alpha, Q_\beta]=0$ for every pair of charges $Q_\alpha, Q_\beta$, and the set of conserved charges also includes the Hamiltonian $H$.
This setting applies to integrable models, as well as models with only a finite number of conserved quantities.

We will now follow Refs.~\cite{Bargheer_2008,Bargheer_2009,Pozsgay2020} and consider deformations of the charges that depend smoothly on a parameter $\lambda$. We define a set of conserved charges $\{Q_\alpha(\lambda)\}$ (and their charge densities $\{q_{\alpha,j}(\lambda)\}$), that for $\lambda=0$ coincides with the original set of (undeformed) charges.
The deformed charges are generated by an operator $X(\lambda)$:
\begin{equation}
\label{eq:deform}
\frac{\mathrm{d} Q_\alpha(\lambda)}{\mathrm{d} \lambda}=i[X(\lambda), Q_\alpha (\lambda)].
\end{equation}
With this definition, $\{Q_\alpha(\lambda)\}$ form a set of mutually commuting charges for any $\lambda$, as can be proven by noting that
\begin{equation}
    \frac{\mathrm{d} [Q_\alpha(\lambda),Q_\beta(\lambda)]}{\mathrm{d} \lambda}=i[X(\lambda), [Q_\alpha (\lambda), Q_\beta(\lambda)]]
\end{equation}
and the initial condition $[Q_\alpha (0), Q_\beta(0)] = 0$.

There are various types of operator $X(\lambda)$ that lead to quasi-local deformations.
This is clearly the case, for example, when the operator $X(\lambda)$ is a local or quasi-local extensive operator.
However, more unconventional classes of operators can also generate quasi-local deformations.
We now consider three classes of operators that generate quasilocal transformations for a generic set of conserved charges.

\subsection{Local extensive operators}
We can consider arbitrary translationally-invariant operators of the form
\begin{equation}
\label{eq:Xex}
X_\text{ex}(\lambda) = \sum_s e_s(\lambda) R_s, \quad R_s = \sum_j r_{s,j} ~,
\end{equation}
where $s$ labels different such operators $R_s$ that we may want to consider (e.g., different types of Pauli strings up to some range), and $e_s(\lambda)$ can be arbitrary functions of $\lambda$.

\subsection{Boosted operators}
One example of non-trivial generators is the class of boosted operators: we consider an operator $X(\lambda)$ of the form
\begin{equation}
\label{eq:Xbo}
X_\text{bo}(\lambda) = -\sum_\beta f_\beta(\lambda) {\mathcal B}[Q_\beta (\lambda)],
\end{equation}
where $f_\beta(\lambda)$ is a real function and $\mathcal B [Q_\beta (\lambda) ]$ is the boosted operator of the charge $Q_\beta (\lambda)$, defined as
\begin{equation}
\mathcal B [Q_\beta(\lambda)]=\sum_j j q_{\beta,j}(\lambda).
\end{equation}

While $\mathcal B [Q_\beta (\lambda)]$ is not an extensive operator, it can be proven that $-i[\mathcal B [Q_\beta (\lambda)],Q_\alpha (\lambda)]$ is extensive when $[Q_\alpha (\lambda), Q_\beta (\lambda)]=0$ is satisfied \cite{Pozsgay2020}: 
because of the commutation relations, we can define the generalized currents  $J_{\beta\alpha,j}(\lambda)$ such that
\begin{equation}
\label{eq:current}
i[Q_\alpha(\lambda), q_{\beta,j}(\lambda)]=J_{\beta\alpha, j}(\lambda)-J_{\beta\alpha, j+1}(\lambda),
\end{equation}
from which we get
\begin{equation}
-i[\mathcal B [Q_\beta (\lambda)],Q_\alpha (\lambda)]=\sum_j J_{\beta\alpha,j}(\lambda) \equiv J_{\beta\alpha; \text{tot}}(\lambda). \label{eq:Jbatot}
\end{equation}
The deformed charges and the generalized currents are expected to be quasi-local in a finite range of $\lambda$ close to $\lambda=0$.

It is important to note that the charge densities (and hence the current operators) are not uniquely defined.
A charge density $\tilde q_{\alpha, j}=q_{\alpha, j}+o_{\alpha,j+1}-o_{\alpha,j}$ corresponds to the same extensive operator $Q_\alpha$.
The corresponding boosted operator then equals the original boosted operator shifted by an extensive local operator:
\begin{equation}
\tilde{X}_\text{bo}(\lambda) = X_\text{bo}(\lambda) + \sum_\beta f_\beta(\lambda) \sum_{j} o_{\beta,j}(\lambda).
\end{equation}
Thus, the non-uniqueness of the definition of the boost of charge operators when defining $X_{\text{bo}}(\lambda)$ is equivalent to allowing adding an arbitrary extensive local operator to $X(\lambda)$.
In the following, we will make a specific choice when defining the boosted operators and will also make the above allowance where needed.

\subsection{Bilocal operators}
Another example of non-trivial generators of quasi-local deformations is the class of bilocal operators, with $X(\lambda)$ defined as
\begin{equation}
\label{eq:Xbi}
X_\text{bi}(\lambda)=\sum_{\beta, \gamma} g_{\beta\gamma}(\lambda)[Q_\beta(\lambda)|Q_\gamma (\lambda)] 
\end{equation}
with 
\begin{multline}
[Q_\beta(\lambda)|Q_\gamma (\lambda)] =\\
=\sum_{j<k} \{q_{\beta,j}(\lambda),q_{\gamma,k}(\lambda) \}+\frac{1}{2}\sum_{j} \{q_{\beta,j}(\lambda), q_{\gamma,j}(\lambda)\}.
\end{multline}
Using Eq. (\ref{eq:current}) we get 
\begin{multline}
\label{eq:bilocal}
i\big[[Q_\beta(\lambda)|Q_\gamma (\lambda)] , Q_\alpha(\lambda)\big]=\\
=\frac{1}{2}\sum_j\{q_{\gamma,j}(\lambda), J_{\beta\alpha,j}(\lambda)+J_{\beta\alpha,j+1}(\lambda)\}\\
-\frac{1}{2}\sum_j\{q_{\beta,j}(\lambda), J_{\gamma\alpha,j}(\lambda)+J_{\gamma\alpha,j+1}(\lambda)\}.
\end{multline}

We note that similarly to the boosted operators case, the non-uniqueness in the definition of the charge density of the form $\tilde{q}_{\beta,j} = q_{\beta,j} + o_{\beta,j+1} - o_{\beta,j}$ leads to the shift of $X_{\text{bi}}$ by an extensive local operator.
On the other hand, shifting $\tilde{q}_{\beta,j}$ by a constant corresponds essentially to adding a boosted operator to the bilocal operator;
equivalently, using a trivially conserved quantity -- the identity operator -- as one of the operators in the bilocal construction gives $[\sum_j 1 | Q_\gamma(\lambda)] = 2  \mathcal{B}[Q_\gamma(\lambda)] + \text{const} \times Q_\gamma(\lambda)$.
In what follows, we are assuming making specific choices for densities and currents when defining the bilocal operators, while the non-uniqueness is taken care of since we are separately including local extensive and boosted operators as generators.

\subsection{Generic deformation}
In general we can define
\begin{equation}
X(\lambda) =X_\text{ex}(\lambda)+X_\text{bo}(\lambda)+X_\text{bi}(\lambda).
\end{equation}
Given the deformed charges $\{Q_\alpha (\lambda)\}$, we can construct a family of Hamiltonians that commute with them, of the form
\begin{equation}
\label{eq:H}
H(\lambda)=\sum_\alpha c_\alpha(\lambda) Q_\alpha(\lambda).
\end{equation}
The coefficients $c_\alpha (\lambda)$ are chosen such that $H(\lambda=0)$ coincides with the original Hamiltonian $H$, but otherwise they can be arbitrary functions of $\lambda$.

\section{Finite order}
\label{sec:finite}
The construction above allows to define deformations of local Hamiltonians that maintain the presence of a set of conservation laws. However, the deformed charges and Hamiltonians obtained with this procedure are not local, but rather quasi-local: they contain arbitrary long-range contributions, with amplitudes that decrease exponentially with the range. In this Section we show how to construct families of Hamiltonians that have strictly finite-range deformed charges that are {\it quasi-conserved}. More precisely, for any integer $\ell>1$, the quasi-conserved charges $Q_\alpha^{<\ell}(\lambda)$ and Hamiltonian $H^{<\ell}(\lambda)$ satisfy
\begin{align}
\label{eq:quasicons1}
[Q_\alpha^{<\ell}(\lambda),Q_\beta^{<\ell}(\lambda)]&=O(\lambda^\ell),\\
\label{eq:quasicons2}
[Q_\alpha^{<\ell}(\lambda),H^{<\ell}(\lambda)]&=O(\lambda^\ell).
\end{align}

To define the quasi-conserved charges, we expand the deformed charges $Q_\alpha(\lambda)$ defined by Eq.~(\ref{eq:deform}) as power series in the small parameter $\lambda$:
\begin{equation}
Q_\alpha (\lambda)=\sum_{n=0}^\infty \frac{\lambda^n}{n!}Q_\alpha^{(n)}.
\end{equation}
Similarly, we expand the functions $e_s(\lambda)$, $f_\beta(\lambda)$, $g_{\beta \gamma}(\lambda)$, and $c_\alpha(\lambda)$, yielding
\begin{equation}
e_s(\lambda)=\sum_{n=0}^\infty \frac{\lambda^n}{n!} e_s^{(n)}, \qquad 
f_\beta (\lambda)=\sum_{n=0}^\infty \frac{\lambda^n}{n!}f_\beta^{(n)},
\end{equation}
\begin{equation}
g_{\beta\gamma} (\lambda)=\sum_{n=0}^\infty \frac{\lambda^n}{n!}g_{\beta\gamma}^{(n)},\qquad 
c_\alpha (\lambda)=\sum_{n=0}^\infty \frac{\lambda^n}{n!}c_\alpha^{(n)}.
\end{equation}
Substituting Eqs.~(\ref{eq:Xex}), (\ref{eq:Xbo}), and (\ref{eq:Xbi}) in Eq.~(\ref{eq:deform}) and equating the two sides order by order in $\lambda$ we get
\begin{align}
&Q_{\alpha}^{(k+1)} =
i\!\!\sum_{m,n=0}^k \frac{k!}{m!n!} \delta_{m+n,k} \sum_s e_s^{(m)} [R_s, Q_\alpha^{(n)}]
\label{eq:Qalpha_kp1} \\
&-i\!\!\!\sum_{m,n,p=0}^k \frac{k!}{m!n!p!}\delta_{m+n+p,k}\sum_\beta f_\beta^{(m)}[\mathcal B [Q_\beta^{(n)}], Q_\alpha^{(p)}] \nonumber \\ 
&+i\!\!\!\!\sum_{m,n,p,t=0}^k \frac{k!}{m!n!p!t!}\delta_{m+n+p+t,k}\sum_{\beta,\gamma} g_{\beta\gamma}^{(m)}[[Q_\beta^{(n)}|Q_\gamma^{(p)}], Q_\alpha^{(t)}] \nonumber
\end{align}
for $k=0,1,2,\dots$.

The above equation shows how to construct the deformed charges order by order in $\lambda$, given the real numbers $e_s^{(n)}$, $f_\beta^{(n)}$, $g_{\beta \gamma}^{(n)}$ that parameterize the deformation.

The quasi-conserved charges $Q_\alpha^{<\ell}(\lambda)$ can then be defined as the truncation of the quasi-local charges to a finite order:
\begin{equation}
Q_\alpha^{<\ell}(\lambda)=\sum_{n=0}^{\ell-1} \frac{\lambda^n}{n!}Q_\alpha^{(n)}.
\end{equation}
Since these differ from $Q_\alpha(\lambda)$ by $O(\lambda^\ell)$, they indeed satisfy Eq.~(\ref{eq:quasicons1}).

Similarly, the Hamiltonian $H^{<\ell}(\lambda)$ satisfying Eq.~(\ref{eq:quasicons2}) can be defined as a truncation of $H(\lambda)$ in Eq.~(\ref{eq:H}):
\begin{equation}
\label{eq:Hquasi}
H^{<\ell}(\lambda)=\sum_{n=0}^{\ell-1} \frac{\lambda^n}{n!}H^{(n)}
\end{equation}
with
\begin{equation}
\label{eq:Hn}
H^{(k)} = \sum_{m,n=0}^k
\frac{k!}{m!n!} \delta_{m+n,k} \sum_\alpha c_\alpha^{(m)} Q_\alpha^{(n)}.
\end{equation}

The existence of a set of quasi-conserved charges has practical consequences for the dynamics generated by the Hamiltonian $H^{<\ell}(\lambda)$, which we discuss in Sec.~\ref{sec:therm}.

\subsection{First order}
We now focus on the case $\ell=2$ and show how to construct the quasi-IoMs $Q_\alpha^{<2}(\lambda) = Q_\alpha^{(0)} + \lambda Q_\alpha^{(1)}$ and the perturbed Hamiltonian $H^{<2}(\lambda) = H^{(0)} + \lambda H^{(1)}$.
The leading correction to the charge is particularly simple:
\begin{align}
\label{eq:k1}
& Q_{\alpha}^{(1)} =
i \sum_s e_s^{(0)} [R_s, Q_\alpha^{(0)}] \\
&-i\sum_\beta f_\beta^{(0)}[\mathcal B [Q_\beta^{(0)}], Q_\alpha^{(0)}] 
+i\sum_{\beta,\gamma} g_{\beta\gamma}^{(0)}[[Q_\beta^{(0)}|Q_\gamma^{(0)}], Q_\alpha^{(0)}], \nonumber
\end{align}
and involves only ``data'' of the unperturbed IoMs $Q_\beta^{(0)}$.

In many examples we consider, the unperturbed Hamiltonian is commonly used to define the charge $Q_2^{(0)}$; specializing to $c_\alpha^{(0)} = c_2^{(0)} \delta_{\alpha,2}$ with some fixed number $c_2^{(0)}$, we have to leading order
\begin{eqnarray}
H^{(0)} &=& c_2^{(0)} Q_2^{(0)} ~, \\
H^{(1)} &=& c_2^{(0)} Q_2^{(1)} + \sum_\alpha c_\alpha^{(1)} Q_\alpha^{(0)} ~.
\label{eq:H1}
\end{eqnarray}

From Eqs.~(\ref{eq:k1}) and (\ref{eq:H1}), we see that we can construct a space of possible ``weak-perturbations'' to first order in $\lambda$ as a linear space spanned by the unperturbed charges $Q_\alpha^{(0)}$, the operators $i[R_s, Q_2^{(0)}]$ (where $R_s$ is a generic extensive operator), the boosted-generated deformations $-i[\mathcal B [Q_\beta^{(0)}], Q_2^{(0)}] = J_{\beta,2;\text{tot}} \equiv J_{\beta;\text{tot}}$ (i.e., the summed current associated to the charge $Q_\beta^{(0)}$), and the bilocal-generated deformations $i[[Q_\beta^{(0)}|Q_\gamma^{(0)}], Q_2^{(0)}]$ [computed as in Eq. (\ref{eq:bilocal})]. 
The fact that current operators are ``nearly integrable'' perturbations was recently noted by Durnin {\it et.~al.}~\cite{Durnin2021} by studying the non-equilibrium dynamics of charges.
Our approach extends this class beyond current operators.
Note, however, that we do not know if this exhausts the full space of all weak-integrability-breaking perturbations to first order---this is an interesting open question.

\section{Examples: Deformations using boosted operators}
\label{sec:ex_boost}
\subsection{Heisenberg chain}
\label{subsec:Heisenberg}
As a first example, we here discuss a particular quasi-integrable deformation of the spin-$1/2$ Heisenberg chain. The undeformed Hamiltonian has the form
\begin{equation}
H^{(0)}=\sum_j \vec \sigma_j\cdot \vec\sigma_{j+1},
\label{eq:H0Heis}
\end{equation}
where $\vec \sigma_j=(\sigma_j^x,\sigma_j^y,\sigma_j^z)$ is the vector of Pauli operators on site $j$. The model is integrable, and the densities of the first few (i.e., with smallest range) conserved charges have the form
\begin{align}
q_{ 2, j}^{(0)} =& \frac{1}{2} \vec\sigma_{j}\cdot\vec\sigma_{j+1} ~, \\
q_{ 3, j}^{(0)} =& -\frac{1}{2} (\vec\sigma_{j}\times\vec\sigma_{j+1})\cdot\vec\sigma_{j+2} ~, \\
q_{ 4, j}^{(0)} =& [(\vec\sigma_j\times \vec \sigma_{j+1})\times \vec \sigma_{j+2}]\cdot \vec \sigma_{j+3} + \nonumber \\
& + \vec\sigma_{j}\cdot\vec\sigma_{j+2}-2 \vec\sigma_{j}\cdot\vec\sigma_{j+1} ~. 
\end{align}
The Hamiltonian is included as $H^{(0)} = 2 Q_2^{(0)}$.
A systematic way of obtaining the higher conserved charges in the Heisenberg chain is by applying the commutator with the boosted operator $\mathcal B[Q_2^{(0)}]$:
\begin{equation}
\label{eq:Qa+1}
    Q_{\alpha+1}^{(0)}=i[\mathcal B[Q_2^{(0)}], Q_\alpha^{(0)}] ,
\end{equation}
see Refs.~\cite{Grabowski1994,Pozsgay2020} and footnote
\footnote{Note that we are using conventions in Pozsgay\cite{Pozsgay2020} and the relation to conventions in Grabovski and Matthiew\cite{Grabowski1994} is $Q_{\alpha}^{(0)} = \frac{(-1)^\alpha}{2} Q_{\alpha}^{(0), \text{GM}}$}.
Appendix~\ref{app:quasiIoMsHeis} shows densities of two more conserved charges obtained this way.

We now want to construct an example of a simple deformation of the Heisenberg Hamiltonian that is generated by a boosted operator and is quasi-integrable to order $\ell=2$.  We therefore need to define the parameters of the deformation $f_\beta^{(0)}$: note that these are the only parameters that define the deformation of the charges to first order in $\lambda$, Eq.~(\ref{eq:k1}), since we are focusing on deformations generated by boosted operators only. %

We consider
\begin{equation}
\label{eq:Q1Heis}
f_\beta^{(0)}=\delta_{\beta,3} ~~\implies~~ Q_\alpha^{(1)}=-i[\mathcal B [Q_3^{(0)}], Q_\alpha^{(0)}],
\end{equation}
and for the first two charges we obtain
\begin{align}
q_{2, j}^{(1)} =& - \frac{1}{2}[(\vec\sigma_j\times \vec \sigma_{j+1})\times \vec \sigma_{j+2}] \cdot \vec \sigma_{j+3} - \vec\sigma_{j} \cdot \vec\sigma_{j+1} ~, \label{eq:q21Heis} \\
q_{3, j}^{(1)} =& \{[(\vec\sigma_j\times \vec \sigma_{j+1})\times \vec \sigma_{j+2}]\times \vec \sigma_{j+3}\}\cdot \vec \sigma_{j+4} \, + \, \label{eq:q31Heis} \\
& + \frac{1}{2} (\vec\sigma_{j}\times\vec\sigma_{j+1} + \vec\sigma_{j}\times\vec\sigma_{j+2})\cdot\vec\sigma_{j+3} ~. \nonumber
\end{align}

We can now define the deformed Hamiltonian using Eqs.~(\ref{eq:Hquasi}) and (\ref{eq:Hn}). Since $H^{(0)}=2Q_2^{(0)}$ we have $c_\alpha^{(0)}=2\delta_{\alpha,2}$ and
\begin{equation}
\label{eq:H1_Heis}
H^{(1)} = 2Q_{2}^{(1)} + \sum_{\alpha}c_\alpha^{(1)}Q_\alpha^{(0)}.
\end{equation}
The Hamiltonian $H^{<2}(\lambda)=H^{(0)}+\lambda H^{(1)}$ is a quasi-integrable deformation of the Heisenberg chain to order $\ell=2$ for any choice of coefficients $c_\alpha^{(1)}$. A particularly relevant model is obtained for $c_\alpha^{(1)}=8\delta_{\alpha,2}+\delta_{\alpha,4}$, in which case
\begin{equation}
H^{(1)} = 2Q_2^{(1)} + 8Q_2^{(0)} + Q_4^{(0)} = \sum_j\vec \sigma_j \cdot \vec \sigma_{j+2}.
\label{eq:H1Heis2ndnb}
\end{equation}
Note that we obtained a range-3 term\footnote{By a range-$m$ term we mean a term that covers $m$ consecutive sites; thus, a nearest-neighbor term is range-2, a second-neighbor term or a three-site term on three consecutive sites is range-3, etc.} by cancelling the range-4 part in $q_2^{(1)}$ by a similar part in $q_4^{(0)}$; this is a general property of the deformation generated by Eq.~(\ref{eq:Q1Heis}) when the unperturbed integrable model contains only nearest-neighbor interactions and its IoMs are obtained using Eq.~(\ref{eq:Qa+1}), see App.~\ref{app:quasiIoMsHeis} for details.
All quantities $Q^{<2}_\alpha = Q_\alpha^{(0)} + \lambda Q_\alpha^{(1)}$ are quasi-IoMs of the deformed model in the sense of Eq.~(\ref{eq:quasicons2}).

The fact that the Heisenberg chain perturbed by the second-neighbor Heisenberg interactions is an example of weak integrability breaking was first noticed in Ref.~\cite{Jung2006} in their calculations of the thermal conductivity of the perturbed Heisenberg chain, and they pointed presence of a quasi-conserved quantity with density proportional to
\begin{equation}
\tilde{q}_{3,j} = q_{3,j}^{(0)} + \lambda \frac{1}{2} (\vec{\sigma}_{j+1} + \vec{\sigma}_{j+2}) \cdot (\vec{\sigma}_j \times \vec{\sigma}_{j+3}) ~,
\label{eq:tildeQ3}
\end{equation}
where we use our convention for writing the leading term.
In App.~\ref{app:quasiIoMsHeis} we show that this quasi-IoM is directly related to $q_3^{<2} = q_3^{(0)} + \lambda q_3^{(1)}$ by adding a combination of the unperturbed conserved densities multiplied by $\lambda$, namely
\begin{equation}
\tilde{q}_{3,j} = q_{3,j}^{<2} + \lambda \left[6 q_{3,j}^{(0)} + \frac{1}{3} q_{5,j}^{(0)} \right] ~,
\label{eq:tildeQ3our}
\end{equation}
which indeed maintains the quasi-conservation property Eq.~(\ref{eq:quasicons2}).

Later work Ref.~\cite{Kurlov2022} used brute-force search for quasi-conserved quantities in this model and found the same quasi-IoM, Eq.~(\ref{eq:tildeQ3}), as well as several other longer-ranged quasi-IoMs, and in App.~\ref{app:quasiIoMsHeis} we show that the next quasi-IoMs are also reproduced by our approach.
We thus suggest that the specific unitary transformation explains all these results; our approach proves that there are in fact as many quasi-IoMs as in the original integrable model and provides a straightforward recipe for obtaining all of them, without the need for brute-force searches.

\subsection{XYZ and XXZ Chain}
\label{subsec:XYZ}
A generalization of the Heisenberg chain is represented by the XYZ chain, with Hamiltonian
\begin{equation}
H^{(0)} = \sum_j (t_x\sigma_j^x\sigma_{j+1}^x + t_y\sigma_j^y\sigma_{j+1}^y + t_z\sigma_j^z\sigma_{j+1}^z) ~.
\end{equation}
Similarly to the case of the Heisenberg chain, the set of conserved charges can be obtained by defining $H^{(0)}=2Q_2^{(0)}$ and using Eq.~(\ref{eq:Qa+1}).
The first two charge densities have the form
\begin{equation}
q_{2,j}^{(0)}=\frac{1}{2}\sum_{\alpha \in \{x,y,z\}}t_\alpha \sigma_j^\alpha \sigma_{j+1}^\alpha ~,
\end{equation}
\begin{equation}
q_{3,j}^{(0)} = -\frac{1}{2} \sum_{\alpha,\beta,\gamma \in \{x,y,z\}} \epsilon_{\alpha\beta\gamma} t_\alpha t_\gamma \sigma_j^\alpha \sigma_{j+1}^\beta \sigma_{j+2}^\gamma ~.
\end{equation}
Expressions for further conserved charges in the XYZ chain can be found, e.g., in Ref.~\cite{Grabowski1995} (with proper translation from their boost definition to ours).
Explicit expressions were also calculated in Ref.~\cite{Nozawa2020}, and in Ref.~\cite{Nienhuis_2021} for the case of the XXZ chain.

We again consider a deformation generated by a boost operator, with  $f_\beta^{(0)} = \delta_{\beta,3}$, and compute the deformed charges $Q_{\alpha}^{(1)}$ to first order using Eq.~(\ref{eq:Q1Heis}); we list the expressions for $Q^{(1)}_2$ and $Q^{(1)}_3$ in App.~\ref{app:XYZ}. 
Using Eq.~(\ref{eq:H1}), the sum of $Q_2^{(1)}$ with any linear combination of the unperturbed charges is then a generic quasi-integrable perturbation to order $\ell=2$.

A possible choice of coefficients $c_\alpha^{(1)}$ of the linear combination that gives a simple (e.g., without three-body terms) perturbation is $c_\alpha^{(1)} = \delta_{\alpha,4}$,
in which case, using appropriate expression for $Q_4^{(0)}$, we get
\begin{align}
H^{(1)} &= 2 Q_2^{(1)} + Q_4^{(0)} \label{eq:XYZH1} \\
&= \sum_j \Big[t_xt_yt_z\vec\sigma_j\cdot\vec\sigma_{j+2}-2\sum_\alpha t_\alpha (\vec{t}^{\;2} - t^2_\alpha)\sigma_j^\alpha\sigma_{j+1}^\alpha \Big] ~,  \nonumber
\end{align}
where we have defined $\vec{t}^{\;2} \equiv \sum_{\alpha \in \{x,y,z\}} t_\alpha^2$.
The reason the specific combination cancels range-4 terms so that only range-3 terms remain is the same as in the Heisenberg case, see App.~\ref{app:detailsBQ3}.

We can readily write the corresponding quasi-conserved quantity $q_{3,j}^{<2} = q_{3,j}^{(0)} + \lambda q_{3,j}^{(1)}$.
We have verified that we can eliminate range 5 terms by combining with $q_{5,j}^{(0)}$ as in the Heisenberg case, Eq.~(\ref{eq:tildeQ3our}). 
Specifically, we find

\begin{align}
& q_{3,j}^{(1)} + \frac{1}{3} q_{5,j}^{(0)} = \!\sum_{\alpha,\beta,\gamma}\! \epsilon_{\alpha\beta\gamma} \left[t_\alpha t_\gamma\left(\frac{2}{3}\vec{t}^{\;2} + t_\beta^2\right) \sigma_j^\alpha \sigma_{j+1}^\beta \sigma_{j+2}^\gamma \nonumber \right. \\
& \left. ~~ - \frac{1}{2} t_\alpha^2 t_\beta t_\gamma \,
\sigma_j^\alpha \sigma_{j+1}^\beta \sigma_{j+3}^\gamma - \frac{1}{2} t_\alpha t_\beta t_\gamma^2 \, \sigma_j^\alpha \sigma_{j+2}^\beta \sigma_{j+3}^\gamma \right].
\end{align}

Note that unlike the Heisenberg case, combining with $q_{3,j}^{(0)}$ does not fully eliminate the first term involving three consecutive sites, so we do not show such combinations.

A particularly relevant case is the XXZ chain, obtained when $t_x=t_y$. To simplify the notation, we set $t_x=t_y=1$, so the unperturbed XXZ Hamiltonian reads
\begin{equation}
H^{(0)} = \sum_j (\sigma_j^x\sigma_{j+1}^x+\sigma_j^y\sigma_{j+1}^y+t_z\sigma_j^z\sigma_{j+1}^z).
\end{equation}
In this case, we can obtain a simpler perturbation with the choice of coefficients
$c_\alpha^{(1)} = \delta_{\alpha,4} + 4 (1+t_z^2) \delta_{\alpha,2}$, which yields
\begin{equation}
H^{(1)} = \sum_j t_z[\vec\sigma_j \cdot \vec\sigma_{j+2} + 2(t_z^2-1)\sigma_j^z\sigma_{j+1}^z].
\label{eq:XXZH1}
\end{equation}
The nearest neighbor term $\sigma_j^z\sigma_{j+1}^z$ simply gives a correction $O(\lambda)$ to the parameter $t_z$, while the next-nearest neighbour Heisenberg interaction can be regarded as the actual weak integrability-breaking perturbation.

To argue this precisely, we need to be careful that the IoMs of the XXZ chain depend on the anisotropy parameter $t_z$, so we need to properly take into account the above shift of the $t_z$ by $O(\lambda)$ when finding the quasi-IoMs for the pure second-neighbor Heisenberg interaction perturbation
\begin{equation}
\tilde{H}^{(1)} \equiv \sum_j \vec{\sigma}_j \cdot \vec{\sigma}_{j+2} ~.
\label{eq:XXZtildeH1}
\end{equation}
To accomplish this, we start with the expected commutation for the found quasi-IoMs for the perturbation $H^{(1)}$:
\begin{equation}
[H^{(0)}(t_z) + \lambda H^{(1)}(t_z), Q_\alpha^{(0)}(t_z) + \lambda Q_\alpha^{(1)}(t_z)] = O(\lambda^2) ~, 
\end{equation}
where we have explicitly indicated that these operators have the $t_z$ parameter in them.
Next, we write
\begin{align*}
& H^{(1)} = g(t_z) \tilde{H}^{(1)} + r(t_z) \sum_j \sigma_j^z \sigma_{j+1}^z \equiv H^{(1)}(t_z) ~, \\
& g(t_z) \equiv t_z ~, \quad r(t_z) \equiv 2 t_z (t_z^2 -1) ~,
\end{align*}
\begin{align*}
& H^{(0)}(t_z) + \lambda H^{(1)}(t_z) = H^{(0)}[t_z + \lambda r(t_z)] + \lambda g(t_z) \tilde{H}^{(1)} \\
& \qquad = H^{(0)}(t_z^\prime) + \lambda g[t_z^\prime - \lambda r(t_z)] \tilde{H}^{(1)} \\
& \qquad = H^{(0)}(t_z^\prime) + \lambda g(t_z^\prime) \tilde{H}^{(1)} + O(\lambda^2) ~,
\end{align*}
where we introduced the parameter
\begin{equation*}
t_z^\prime \equiv t_z + \lambda r(t_z) \leftrightarrow t_z = t_z^\prime - \lambda r(t_z) = t_z^\prime - \lambda r(t_z^\prime) + O(\lambda^2) ~,
\end{equation*}
and in the end expanded to the exhibited order in $\lambda$.
Repeating the same for the quasi-IoM parts and plugging into the above commutator, we obtain
\begin{align*}
& \Bigg[H^{(0)}(t_z^\prime) + \lambda g(t_z^\prime) \tilde{H}^{(1)}, \\
& ~~ Q_\alpha^{(0)}(t_z^\prime) + 
\lambda \left(Q_\alpha^{(1)}(t_z^\prime) - r(t_z^\prime) \frac{\partial Q_\alpha^{(0)}(t_z^\prime)}{\partial t_z^\prime} \right)\Bigg] = O(\lambda^2) ~.
\end{align*}
At this point we can drop the prime on the parameter $t_z^\prime$ and conclude that indeed $\tilde{H}^{(1)}$ is a weak integrability breaking perturbation for the XXZ chain at any value of the anisotropy parameter and also read off the corresponding quasi-IoMs in terms of the ones obtained for $H^{(1)}$ above.

We note that this isotropic next-nearest-neighbor perturbation of the XXZ chain was considered in \cite{Jung2006,Jung2007}, where slow relaxation was signalled by a heat conductivity of order $\sim\lambda^{-4}$ (much larger than the generically expected scaling $\sim\lambda^{-2}$), and explained as the consequence of the existence of a quasi-conserved quantity.
Here we found an explicit form of the quasi-IoM derived from $Q_{3}^{(0)}$.
Furthermore, our approach shows that there is, in fact, an extensive number of such quasi-conserved quantities.

The above demonstration of a near-integrability of $\tilde{H}^{(1)}$ [Eq.~(\ref{eq:XXZtildeH1})] starting from the near-integrability of $H^{(1)}$ [Eq.~(\ref{eq:XXZH1}), derived from the truncated deformations], in fact applies quite generally for integrable models with a continuosuly varying parameter.
Specifically, suppose the unperturbed Hamiltonian has the form 
\begin{equation}
H^{(0)}(u) = H_{\text{I}}^{(0)} + u H_{\text{II}}^{(0)} 
\end{equation}
with a parameter $u$ appearing as a coefficient of some part of the Hamiltonian, where we assume that both $H_{\text{I}}^{(0)}$ and $H_{\text{II}}^{(0)}$ have no $u$ dependence in them (they may depend on some other parameters, which however are kept fixed throughout).
Suppose that $H^{(0)}(u)$ has IoMs $Q_\alpha^{(0)}(u)$, which in general depend on $u$ (and in more complicated ways than $H^{(0)}$).
Then $H^{(1)} = H_{\text{II}}$ can be considered as a weak (nearly-integrable) perturbation of $H^{(0)}(u)$ at fixed $u$,
with $H = H^{(0)}(u) + \lambda H_{\text{II}}$ having the quasi-IoMs $Q^{<\ell=2}_\alpha = Q_\alpha^{(0)}(u) + \lambda \partial_u Q_\alpha^{(0)}(u)$, which is simply the first term in the Taylor expansion of $Q_\alpha^{(0)}(u+\lambda)$.
Of course, by continuing the Taylor expansion we can write quasi-IoMs for the specific perturbation to arbitrary order $\ell$, but our main interest is $\ell=2$ where we can combine the above $H^{(1)}$ with general nearly-integrable perturbations since these form a linear space:
If $H^{(1)\prime}$ and $H^{(1)\prime\prime}$ are nearly-integrable perturbations with the quasi-IoM corrections $Q_\alpha^{(1)\prime}$ and $Q_\alpha^{(1)\prime\prime}$ respectively, then $a' H^{(1)\prime} + a'' H^{(1)\prime\prime}$ is also a nearly-integrable perturbation with the quasi-IoM correction $a' Q_\alpha^{(1)\prime} + a'' Q_\alpha^{(1)\prime\prime}$, for arbitrary $a'$ and $a''$.

As an immediate application, the unperturbed XYZ model has parameters $t_x$, $t_y$, and $t_z$, and hence we can add independent nearest neighbor terms $\sigma_j^x \sigma_{j+1}^x$, $\sigma_j^y \sigma_{j+1}^y$, and $\sigma_j^z \sigma_{j+1}^z$ while preserving $\ell=2$ near-integrability.
In this way, starting with Eq.~(\ref{eq:XYZH1}) we conclude that the second-neighbor Heisenberg interaction is a nearly integrable perturbation for the XYZ chain with arbitrary anisotropies (with correspondingly recalculated quasi-IoMs).

\subsection{Hubbard model}
\label{subsec:Hubbard}
The Hamiltonian of the Hubbard model reads
\begin{eqnarray}
H^{(0)} &=& -2\sum_{j,s=\uparrow,\downarrow} (a_{j,s}^\dagger a_{j+1,s}+a^\dagger_{j+1,s}a_{j,s})\nonumber\\
&+& 4U\sum_j \left(n_{j,\uparrow}-\frac{1}{2}\right)\left(n_{j,\downarrow}-\frac{1}{2}\right).
\end{eqnarray}
We use notation from Ref.~\cite{Grabowski1995} for easy referencing to their expressions for the IoMs.
We define $Q_{2}^{(0)}=H^{(0)}$. 
In contrast to the Heisenberg, XYZ, and XXZ chains, where the conserved charges can be obtained using $\mathcal{B}[Q_2^{(0)}]$ as a ``ladder'' operator [Eq.~(\ref{eq:Qa+1})], no similar systematic construction can be used to find the conserved charges in the Hubbard model, and brute force methods have been used instead \cite{Grabowski1995,fukai2023all}. 
The first IoM is:
\begin{align}
Q_3^{(0)} &= -2i\sum_{j,s} (a_{j,s}^\dagger a_{j+2,s} - \text{H.c.}) \\
& + 4 i U \sum_{j,s} (a_{j,s}^\dagger a_{j+1,s} - \text{H.c.})(n_{j,-s} + n_{j+1,-s} - 1) ~, \nonumber
\end{align}
where ``$-s$'' denotes the opposite spin to $s$, i.e., $-s = \downarrow,\uparrow$ for $s = \uparrow,\downarrow$.
This IoM is symmetric (even) under the physical spin SU(2) and pseudo-spin (a.k.a.\ $\eta$-pairing) SU(2) symmetries of the Hubbard chain but is odd under the time reversal and inversion symmetries.
In what follows, we define the densities $q_{2,j}^{(0)}$ and $q_{3,j}^{(0)}$ such that they are respectively even and odd under the inversion in the bond center between $j$ and $j+1$, see Eqs.~(\ref{eq:q2Hubb}) and (\ref{eq:q3Hubb}) in  App.~\ref{app:Hubbard}.

Since $\mathcal{B}[Q_2^{(0)}]$ does not generate IoMs, we can consider deformations generated by the boost operator $\mathcal{B}[Q_2^{(0)}]$, as they will correspond to proper perturbations of the model, instead of integrals of motion. We can therefore define
\begin{equation}
Q_\alpha^{(1)}=-i[\mathcal{B}[Q_2^{(0)}], Q_\alpha^{(0)}], \quad H^{(1)}=Q_2^{(1)}+\sum_\alpha c_\alpha^{(1)} Q_{\alpha}^{(0)} .
\end{equation}
Some examples of such first-order weak integrability breaking perturbations $H^{(1)}$ that can be generated in this way are
\begin{equation}
Q_{2}^{(1)} + Q_{3}^{(0)} = 2i \sum_{j,s=\uparrow,\downarrow} (a_{j,s}^\dagger a_{j+2,s} - \text{H.c.}) ~,
\end{equation}
or
\begin{align}
Q_{2}^{(1)} + 2Q_{3}^{(0)} = 4iU\sum_{j,s} & (a_{j,s}^\dagger a_{j+1, s} - \text{H.c.}) \times \nonumber \\
& \times (n_{j,-s} + n_{j+1,-s} - 1) ~.
\end{align}
These terms preserve the spin SU(2) and pseudo-spin SU(2) symmetries but break the time reversal and inversion symmetry.
In particular, we see that the simplest hopping modification of the Hubbard model that preserves both SU(2) symmetries---namely, the second-neighbor pure imaginary hopping---is in fact a weak integrability breaking perturbation.

Note that to obtain a weak perturbation $V = i[X, Q_2^{(0)}]$ that respects both inversion and time reversal symmetry, we need $X$ to be invariant under inversion but odd under time reversal.
Therefore, if $X$ is an extensive local operator, it cannot be a fermion bilinear and, as a consequence, $V$ will contain terms with more than four fermionic operators.
Perturbations with only two and four fermionic operators that respect both the time reversal and inversion symmetries can instead be generated using boosted operators, for example, $\mathcal B[Q_3^{(0)}]$ with $q_{3,j}^{(0)}$ given by Eq.~(\ref{eq:q3Hubb}).
We obtain the corresponding operator, Eq.~(\ref{eq:Jbatot}),
\begin{align}
&J_{3,2; \text{tot}}^{(0)} =
4 \sum_{j,s} (a_{j,s}^\dagger a_{j+1,s} - a_{j,s}^\dagger a_{j+3,s} + \text{H.c.}) \label{eq:HubbJ32tot} \\
&-32U \sum_j \left(n_{j,\uparrow}-\frac{1}{2}\right) \left(n_{j,\downarrow}-\frac{1}{2}\right) \nonumber \\
&+4U \sum_{j,s} \Big[ 2(a_{j,s}^\dagger a_{j+1,s} - \text{H.c.})(a_{j+1,-s}^\dagger a_{j+2,-s} - \text{H.c.}) \nonumber \\
&+(a_{j,s}^\dagger a_{j+2,s} + \text{H.c.}) (n_{j,-s} + 2n_{j+1,-s} + n_{j+2,-s} - 2)\Big]. \nonumber
\end{align}
This perturbation contains fermion hopping up to range 4 and four-fermion interactions up to range 3.
Note also that we can absorb the nearest-neighbor hopping and the on-site Hubbard terms into an $O(\lambda)$ shift of the parameter $U$, and hence the remaining parts of $J_{3,2;\text{tot}}^{(0)}$ can also be viewed as weak integrability breaking perturbations.
Furthermore, we can remove the range-4 term by combining with $Q_4^{(0)}$ listed in App.~\ref{app:Hubbard}, Eq.~(\ref{eq:HubbQ4}), at the expense of introducing additional range-3 terms, including also six-fermion terms.
However, by combining with another weak integrability breaking term generated using $V = i[X, Q_2^{(0)}]$ with an extensive local operator $X$ with only nearest-neighbor terms and that has both SU(2) symmetries and is invariant under the inversion but odd under the time reversal, we can eliminate the six-fermion terms leaving only four-fermion terms up to range 3.
The final result is listed in Eq.~(\ref{eq:Hubbard_simplest_pert}), showcasing how fairly complicated generalized current perturbations can be turned into simpler perturbations using additional freedoms discussed in this paper.
For details, we refer the readers to App.~\ref{app:Hubbard}.

\section{Examples: Deformations using bilocal operators}
\label{sec:ex_biloc}

\subsection{Free spinless fermions}
\label{subsec:freeferms}
As a simple example to illustrate the deformations induced by bilocal operators, we here consider a model of free spinless fermions hopping on a chain:
\begin{equation}
H^{(0)} = -\sum_j (a_j^\dagger a_{j+1} + \text{H.c.}).
\end{equation}
For simplicity, we consider the case of real nearest-neighbor hoppings, but our discussion can be extended to generic complex hoppings of arbitrary ranges.
We define the following set of conserved quantities with densities:
\begin{align}
\nonumber q_{1,j}'^{(0)}&=n_j, & &\\ 
\nonumber
q_{2,j}'^{(0)}&=-(a_j^\dagger a_{j+1}+\text{H.c.}),\quad &
q_{2,j}''^{(0)}&=-i(a_j^\dagger a_{j+1}-\text{H.c.}),\\
\nonumber
\dots & & &\\
q_{m,j}'^{(0)}&=-(a_j^\dagger a_{j+m-1}+\text{H.c.}),&\,  q_{m,j}''^{(0)}&=-i(a_j^\dagger a_{j+m-1}-\text{H.c.}).
\end{align}
We are interested in the family of deformations of the Hamiltonian $H^{(0)}=Q_2'^{(0)}$ that are weak integrability breaking to first order in $\lambda$. 
In particular, we focus on the contributions in $Q_2'^{(1)}$ that correspond to interaction terms. 
While boosted operators of the charges $Q_\alpha'^{(0)}, Q_\alpha''^{(0)}$ can only generate fermion bilinears, bilocal operators can generate interactions: since the conserved quantities and the generalized currents are fermion bilinears, from Eqs.~(\ref{eq:k1}) and (\ref{eq:bilocal}) we see that the linear space of deformations $Q_2'^{(1)}$ generated by bilocal operators contains in general four-fermion operators.
Some examples are:
\begin{align}
(a):&\quad    
i[[Q_2'^{(0)}|Q_1'^{(0)}], Q_2'^{(0)}]=-2\sum_j  q_{3,j}''^{(0)} \cdot n_{j+1},\\
(b):&\quad 
i[[Q_2''^{(0)}|Q_1'^{(0)}], Q_2'^{(0)}]=2\sum_j (q'^{(0)}_{3,j}+2n_j) n_{j+1},\\
(c):&\quad
i[[Q_2'^{(0)}| Q_2''^{(0)}], Q_2'^{(0)}]=-2\sum_j q_{2,j}'^{(0)}\cdot(n_{j-1}+n_{j+2}).
\end{align}
As an illustration,  in App.~\ref{app:freefer} we show the corresponding quasi-conserved quantities in each case obtained by  applying the same deformations to  $Q_3'^{(0)}$. 

Note that there are total of eight linearly-independent translationally invariant and U(1)-charge conserving Hermitian four-fermion terms of range up to 3:
\begin{align}
& \sum_j n_j n_{j+1}; \quad 
\sum_j n_j n_{j+2}; \\
& \sum_j n_j (U a_{j+1}^\dagger a_{j+2} + \text{H.c.}), \quad U \in \mathbb{C}; \\ 
& \sum_j (V a_j^\dagger a_{j+1} + \text{H.c.}) n_{j+2}, \quad V \in \mathbb{C}; \\
& \sum_j n_{j+1} (W a_j^\dagger a_{j+2} + \text{H.c.}), \quad W \in \mathbb{C}.
\end{align}
Thus we see that three of these eight directions in such a space of range-3 perturbations are in fact weak integrability breaking perturbations.
In fact, we know one more weak integrability breaking perturbation of range 3 obtained by using an extensive local operator $X = \sum_j n_j n_{j+1}$ as a generator:
\begin{equation}
(d): \quad i[X, Q_2^{(0)}] = \sum_j  q_{2,j}''^{(0)}(n_{j-1} - n_{j+2}) ~.
\end{equation}
We can further organize these perturbations by their transformation properties under the lattice inversion and time reversal (defined as complex conjugation in the number basis).
Thus, out of the eight terms, there are four terms that are invariant under both the inversion and time reversal, namely the two density-density terms, the combination $U=V \in \mathbb{R}$, and the term $W \in \mathbb{R}$.
The $U=V \in \mathbb{R}$ term is in fact the weak integrability breaking perturbation (c), while the $W \in \mathbb{R}$ term is a combination of the perturbation (b) and the nearest-neighbor density-density term.
Furthermore, adding just the nearest-neighbor density-density interaction in fact leads to another integrable model equivalent to the XXZ chain, and by argument similar to Sec.~\ref{subsec:XYZ} any linear combination of this term and the terms (b) and (c) will also be a weak integrability breaking of the free-fermion chain.
Hence we conclude that among the inversion and time-reversal invariant range-3 perturbations only the second-neighbor density-density interaction truly breaks the integrability in the leading order.

\subsection{Quantum Ising chain}
\label{subsec:QIsing}
The transverse field quantum Ising chain has the following Hamiltonian:
\begin{equation}
H^{(0)} = -J \sum_j(\sigma_j^x \sigma_{j+1}^x + h\sigma_j^z).
\end{equation}
The model can be solved by a Jordan Wigner transformation, which maps the Hamiltonian to a free fermionic model.
We here list the first few conserved quantities, with densities $q_{\alpha,j}^{(0)}$ \cite{Grady1982}:
\begin{align}
q_{2,j}^{(0)}=&\sigma_j^x\sigma_{j+1}^x+\frac{h}{2} (\sigma_j^z+\sigma_{j+1}^z) ~, \\
q_{3,j}^{(0)}=&\sigma_j^y\sigma_{j+1}^x-\sigma_j^x\sigma_{j+1}^y ~, \\
q_{4,j}^{(0)}=&\frac{1}{2}(\sigma_{j-1}^x\sigma_{j}^z\sigma_{j+1}^x+\sigma_j^x\sigma_{j+1}^z\sigma_{j+2}^x-\sigma_j^z-\sigma_{j+1}^z) \nonumber \\
&-h(\sigma_j^x\sigma_{j+1}^x+\sigma_j^y\sigma_{j+1}^y) ~.
\end{align}
Similarly to the case of free spinless fermions, bilocal deformations can be used to generate weak integrability breaking perturbations of the quantum Ising chain that have the form of fermion interactions.
Boosted operators, on the other hand, can only generate free-fermion terms.

As a first example of a perturbation induced by a bilocal  generator, we consider the operator
\begin{equation}
i[[Q_2^{(0)}|Q_3^{(0)}], Q_2^{(0)}] = 4h \sum_j (\sigma_j^x\sigma_{j+2}^x + \sigma_j^z\sigma_{j+1}^z).
\label{eq:IsingBiloc23}
\end{equation}
Dropping the unimportant scale factor, we conclude that $\hat{V} \equiv \sum_j (\sigma_j^x\sigma_{j+2}^x + \sigma_j^z\sigma_{j+1}^z)$ is a weak integrability breaking perturbation.
This perturbation is self-dual under the Kramers-Wannier transformation.
Under the Jordan-Wigner transformation, it gives four-fermion interactions consisting of products of Majoranas on four consecutive sites.
Interestingly, at the Ising critical point $h=1$, the perturbed Hamiltonian $\hat{H_0} + \lambda \hat{V}$ maps precisely to the interacting Majorana chain studied in Ref.~\cite{Rahmani2015}.
It would be interesting to revisit their study in light of our conclusion that this interaction breaks the integrability only in the next order.
For example, one may wonder if this may explain the very large coupling needed to reach the tricritical point, but this requires detailed considerations which we leave for future work.

Another example of a weak perturbation of range $3$ induced by a bilocal deformation is
\begin{align}
\label{eq:Isingbil}
i[[Q_2^{(0)}|Q_4^{(0)}], Q_2^{(0)}] &= 2h\sum_j (-h\sigma_j^x\sigma_{j+2}^y+h\sigma_j^y\sigma_{j+2}^x \nonumber \\
& -\sigma_j^x\sigma_{j+1}^y\sigma_{j+2}^z+\sigma_j^z\sigma_{j+1}^y\sigma_{j+2}^x) ~. 
\end{align}
Under the Ising duality \footnote{Under the Ising duality, the operators $\sigma_j^x \sigma_{j+1}^x$ and $\sigma_j^z$ are mapped to $\tau_{j+1/2}^z$ and $\tau_{j-1/2}^x \tau_{j+1/2}^x$ respectively, where $\tau^{x,y,z}$ are Pauli operators defined on the links of the original lattice.  Thus, the first two operators in the RHS of Eq.~(\ref{eq:Isingbil}) transform as $-\sigma_j^x \sigma_{j+2}^y = -i(\sigma_j^x \sigma_{j+1}^x) (\sigma_{j+1}^x \sigma_{j+2}^x) \sigma_{j+2}^z\rightarrow -i\tau_{j+1/2}^z \tau_{j+3/2}^z (\tau_{j+3/2}^x \tau_{j+5/2}^x) = \tau_{j+1/2}^z \tau_{j+3/2}^y \tau_{j+5/2}^x$, and similarly $\sigma_j^y \sigma_{j+2}^x=-i\sigma_{j}^z(\sigma_j^x\sigma_{j+1}^x)(\sigma_{j+1}^x\sigma_{j+2}^x)\rightarrow -i(\tau_{j-1/2}^x \tau_{j+1/2}^x)\tau_{j+1/2}^z \tau_{j+3/2}^z =-\tau_{j-1/2}^x \tau_{j+1/2}^y \tau_{j+3/2}^z$}, the first Pauli product (including the sign) is interchanged with the fourth one and the second is interchanged with the third.
Thus, this perturbation is self-dual only when $h=1$, which is different from the previous perturbation Eq.~(\ref{eq:IsingBiloc23}).
Furthermore, this perturbation is not invariant under an anti-unitary symmetry of the original Ising model defined as a complex conjugation in the $\sigma^x$ basis, so it is a bit less natural perturbation to consider.

\subsection{Heisenberg chain}
\label{subsec:XbiHeis}
In Secs.~\ref{subsec:Heisenberg} and \ref{subsec:XYZ} we considered weak perturbations of the Heisenberg, XYZ, and XXZ chains generated by boosted operators.
Bilocal operators can be used as generators to obtain more examples of weak perturbations of these models.
Once again, we focus on the perturbations that have the same symmetries as the original Hamiltonian.
For example, with the choice of generator $X(\lambda)=[Q_2(\lambda)|Q_3(\lambda)]$, the deformed charges preserve their parity under inversion and time reversal.
For concreteness, we consider the deformed charges of the Heisenberg chain obtained with this generator. 
The first order deformation $Q_2^{(1)}$ reads:
\begin{multline}
i[[Q_2^{(0)}|Q_3^{(0)}], Q_2^{(0)}] = \sum_j \Big[ \vec \sigma_j \cdot \vec \sigma_{j+2} - \vec \sigma_j \cdot \vec \sigma_{j+3} \\
+\frac{1}{2} (\vec \sigma_j \cdot \vec \sigma_{j+1})(\vec \sigma_{j+2} \cdot \vec \sigma_{j+3}) + \frac{1}{2} (\vec \sigma_j \cdot \vec \sigma_{j+3})(\vec \sigma_{j+1} \cdot \vec \sigma_{j+2} ) \Big].
\end{multline}
This perturbation is manifestly invariant under inversion and time reversal and preserves the SU$(2)$ symmetry of the Heisenberg Hamiltonian.
We conjecture that it is not possible to generate this weak-integrability-breaking perturbation using the boosted IoMs as generators, i.e., it genuinely requires the bilocal operators as generators.

To give some perspective on the above result, we note that the space of range-4 terms that are symmetric under spin-SU(2), lattice translation and inversion, and time reversal has dimension six.
Two directions in this space ($Q_2^{(0)}$ and $Q_4^{(0)}$) are integrable, while the above four-spin term and the second-neighbor Heisenberg interaction considered in Sec.~\ref{subsec:Heisenberg} are independent nearly-integrable directions.
Thus, only two out of six directions truly break the integrability at the leading order, so even including range-4 perturbations the integrability of the Heisenberg chain is more robust than one would naively expect.

\section{Examples: deformations to higher orders}
\label{sec:ex_higher}
In the examples discussed so far we have examined various types of weak perturbations of integrable Hamiltonians of the form
\begin{equation}
H^{<2} = H^{(0)} + \lambda H^{(1)} ~,
\end{equation}
such that a set of quasi-conserved quantities $Q_\alpha^{<2}$ can be defined with the property that $[H^{<2}, Q_\alpha^{<2}] = O(\lambda^2)$, $[Q_\alpha^{<2}, Q_\beta^{<2}] = O(\lambda^2)$.

It is possible to apply the general procedure in Sec.~\ref{sec:finite} to obtain perturbations of the Hamiltonian and of the quasi-conserved charges, such that they commute up to terms of order $\lambda^\ell$ with ${\ell>2}$. 
 Specifically, for $\ell=3$ we consider
\begin{equation}
H^{<3} = H^{(0)} + \lambda H^{(1)} + \frac{\lambda^2}{2}H^{(2)} ~,
\label{eq:Hl3gen}
\end{equation}
where the $H^{(k)}$ are defined as in Eq. (\ref{eq:Hn}):
\begin{eqnarray}
H^{(0)} &=& \sum_\alpha c_\alpha^{(0)} Q_\alpha^{(0)} = c_2^{(0)} Q_2^{(0)} ~, \\
H^{(1)} &=& c_2^{(0)} Q_2^{(1)} + \sum_\alpha c_\alpha^{(1)} Q_\alpha^{(0)} ~, \\
\label{eq:H2}
H^{(2)} &=& c_2^{(0)} Q_2^{(2)} + \sum_\alpha \big(2c_\alpha^{(1)} Q_\alpha^{(1)} + c_\alpha^{(2)} Q_\alpha^{(0)} \big) ~.
\end{eqnarray}
Here we specialized to  $c_\alpha^{(0)} = c_2^{(0)} \delta_{\alpha,2}$ intending to work around the unperturbed Hamiltonian, which in the examples we consider is commonly used to define the charge $Q_2^{(0)}$.

\subsection{Heisenberg chain}
In Sec.~\ref{subsec:Heisenberg} we showed that a weak first-order perturbation of the Heisenberg chain [$H^{(0)}$ in Eq.~(\ref{eq:H0Heis}), with convention $c_2^{(0)}=2$] of the form $H^{(1)} = \sum_j \vec\sigma_j \cdot \vec\sigma_{j+2}$ can be obtained by choosing $f_\beta^{(0)}=\delta_{\beta,3}$ and $c_\alpha^{(1)}=8\delta_{\alpha,2}+\delta_{\alpha,4}$ [cf.\ Eq.~(\ref{eq:H1Heis2ndnb})]. 
We now want to show how to construct $H^{(2)}$ such that $H^{<3}$ has a set of quasi-conserved quantities up to order $\lambda^3$.
From Eq.~(\ref{eq:H2}) we get
\begin{equation}
H^{(2)} = 2Q_2^{(2)} + 16Q_2^{(1)} + 2Q_4^{(1)} + \sum_\alpha c_\alpha^{(2)} Q_\alpha^{(0)} ~,
\label{eq:H22_Heis_ell3}
\end{equation}
where using Eq.~(\ref{eq:Qalpha_kp1}) with $f^{(0)}_\beta = \delta_{\beta,3}$ we have $Q_\alpha^{(1)} = -i[\mathcal B[Q_3^{(0)}], Q_\alpha^{(0)}]$
and
\begin{align}
Q_2^{(2)} &= -i[{\mathcal B}[Q_3^{(0)}], Q_2^{(1)}] - i[{\mathcal B}[Q_3^{(1)}], Q_2^{(0)}] \nonumber \\
& ~~~ -i \sum_\beta f_\beta^{(1)}[{\mathcal B}[Q_\beta^{(0)}], Q_2^{(0)}] ~.
\label{eq:HeisQ22}
\end{align}
With this definition, $H^{<3}$ satisfies the desired property for any choice of coefficients $f_\beta^{(1)}$ and $c_\alpha^{(2)}$. 
We can use this freedom to look for perturbations that have small range and involve a small number of spins.
An example of a particularly simple deformation of the Heisenberg chain that is quasi-integrable to order $\ell=3$ is 
\begin{equation}
H^{<3}(\lambda) = \sum_j \left(\vec\sigma_j\cdot \vec\sigma_{j+1} + \lambda\,\vec\sigma_j \cdot \vec\sigma_{j+2} + \lambda^2 \vec\sigma_j \cdot \vec\sigma_{j+3}\right).
\label{eq:Hl3Heis}
\end{equation}
We refer the readers to App.~\ref{subapp:ell3Heis} for details on the specific choices of coefficients $f_\beta^{(1)}$ and $c_\alpha^{(2)}$ and some intermediate steps.
The degree of simplification that we managed to achieve is quite surprising given the complicated intermediate expressions, and we are wondering if there may be some reason for this.
It would be interesting to check if one can achieve comparable simplification at the next order.

\section{Thermalization time}
\label{sec:therm}
We now discuss the implications of the existence of quasi-conserved quantities for the thermalization time.
In a generic quench (with an arbitrary initial state), the local quantities $Q_\alpha^{<\ell}(\lambda)$ are conserved by the Hamiltonian $H^{<\ell}(\lambda)$ for at least a time $t \sim O(\lambda^{-\ell})$ (see App.~\ref{sec:thermtime}). 
We remark that this lower bound on the thermalization time is a completely rigorous bound using only the locality of the Hamiltonian and of the quasi-conserved observable.

On the other hand, if we use a perturbative calculation for the time-averaged rate of decay of the conserved quantity in the spirit of the Fermi's Golden Rule (see App.~\ref{sec:thermtime} for precise meaning), we would get a (non-rigorous) estimate for the thermalization time as $O(\lambda^{-2\ell})$.
This estimate can be intuitively understood by noting that in our construction the Hamiltonian is $H^{<\ell}(\lambda) = H(\lambda) + O(\lambda^\ell)$, with $H(\lambda)$ integrable, and therefore the effective integrability-breaking perturbation strength is $\lambda^\ell$.
On a time scale $t \ll \lambda^{-\ell}$ the dynamics is determined by $H(\lambda)$, and the system prethermalizes to a generalized Gibbs ensemble with conserved charges $Q_\alpha(\lambda)$.
Genuine thermalization is triggered by the effective perturbation $O(\lambda^\ell)$, whose effect on the dynamics becomes non-negligible at longer times. 
Using standard estimates of the rate, but noting that the perturbation strength is $\lambda^\ell$, we then expect thermalization after time $\tau \sim \lambda^{-2\ell}$.

We note that this argument agrees with the direct perturbative estimates of the rate in the important case $\ell=2$: 
For such special weak perturbations the formal Fermi's Golden Rule rate $O(\lambda^2)$ vanishes after a time $O(1)$ (see App.~\ref{sec:thermtime}).
The vanishing of the $\lambda^2$ order of the rate indicates that after a decay at small times, the expectation value of an observable $Q^{(0)}_\alpha$ (i.e., one of the original charges) reaches a plateau.
Using this observation, it was argued in \cite{Durnin2021}, that these perturbations do not lead to thermalization in the Boltzmann regime (i.e., in the limit $\lambda \to 0$, $t \to \infty$ with $\lambda^2 t = \text{const.}$), but lead to hydrodynamic diffusion. 

The observable $Q_\alpha^{(0)}$ reaching a plateau after $O(1)$ time is consistent with a picture where $Q^{(0)}_\alpha$ has a component onto the quasi-conserved $Q^{<2}_\alpha = Q^{(0)}_\alpha + \lambda Q^{(1)}_\alpha$ that does not thermalize until a much later time.
In fact, it is much easier to see the vanishing $O(\lambda^2)$ rate by thinking directly about the quasi-conserved $Q^{<2}_\alpha$ instead of the original $Q^{(0)}_\alpha$: the formal $O(\lambda^2)$ rate of change of $Q^{<2}_\alpha$ is identically zero at any time (and not just after $O(1)$ time).

In App.~\ref{sec:thermtime} we also consider formal $O(\lambda^3)$ term in the rate of change of $Q^{<2}_\alpha$ and show that it vanishes after $O(1)$ time.
Hence the leading non-vanishing rate after $O(1)$ time is actually $O(\lambda^4)$, in agreement with our picture and the intuition based on the effective perturbation of the integrable model $H(\lambda)$.
The above statements about the rates are for initial states or ensembles generated by the unperturbed integrable model (i.e., defined by $\{ Q_\alpha^{(0)} \}$).
In App.~\ref{sec:thermtime} we also show initial ensembles where the rate of change of the quasi-conserved $Q^{<2}_\alpha$ starts at $O(\lambda^4)$ for \emph{all times}, which can be viewed as formalizing the above intuition by also finding appropriate ``quasi-stationary'' initial states.
Thus we have established a good understanding of the connection between direct perturbation theory calculations in the case of such nearly-integrable perturbations and thinking in terms of the quasi-conserved quantities, observing both the physical intuition and analytical power of the latter framework.

\section{Conclusions}
\label{sec:conclusions}
We have here demonstrated a general construction for producing weak (i.e., nearly integrable) perturbations of integrable lattice models to arbitary order $\lambda^\ell$.
These weakly-perturbed models have an extensive number of (extensive local) approximate conserved quantities that commute with the Hamiltonian up to corrections $O(\lambda^\ell)$.
We have applied the construction to several well-known spin chains and fermionic models, focusing on finding particularly simple such examples and also putting some previously known instances and their physics into a unified framework (e.g., the relation between the vanishing of the Fermi's Golden Rule rate and the presence of the quasi-IoM).

In the case of truncation to linear order, i.e., achieving commutation up to corrections $O(\lambda^2)$, this approach produces a linear space of weak integrability breaking perturbations~(``diffusive subspace''~\cite{Durnin2021}, as opposed to generic ``thermalizing'' perturbations).
While this subspace is of course measure zero in the full space of possible perturbations, it may nevertheless play an important role in realistic physics applications.
For example, for the spin-1/2 Heisenberg chain, it turns out that all SU(2)-spin symmetric perturbations that are translationally invariant and have range up to 3 (i.e., involving up to three consecutive spins) are in fact either integrable or weak integrability breaking perturbations.
As discussed at the end of Sec.~\ref{subsec:XbiHeis},
for range-4 perturbations, if we also require lattice inversion and time reversal symmetry, three out of five perturbations (not counting $Q_2^{(0)}$) of the Heisenberg chain are either integrable or weak integrability breaking.
Such unexpected paucity of natural true integrability breaking perturbations of the Heisenberg chain may help explain the robustness of the superdiffusion signatures in numerical studies~\cite{DeNardis2021} as well as in physical world experiments~\cite{Scheie2021, Wei2022}.

As another example in the same spirit,
for the spinless fermion chain, in the four-dimensional space of inversion and time-reversal symmetric four-fermion interactions up to range 3, three directions are in fact weak integrability breaking perturbations.
In general,  when we study effects of a given perturbation, we would want to understand/remove appropriate ``components'' onto the weak integrability breaking ones, since only the remaining part represents generic integrability breaking perturbation that dominates the thermalization rate of the system.
Furthermore,
in some situations such a removal may not be possible, e.g., under sufficiently restrictive range and symmetry conditions, and it is important to understand the thermalization rates in such cases as well.
This shows importance of systematic constructions of weak integrability breaking perturbations, and we hope that our work will stimulate further such studies. 

A very interesting Ref.~\cite{Pandey2020} proposed to analyze perturbations to integrable and non-integrable systems by constructing so-called Adiabatic Gauge Potential (AGP)~\cite{Kolodrubetz2017} which is an analog of the generator $X(\lambda)$ in our formalism.
While a regularized AGP always exists, for true integrability breaking perturbations it is expected to be highly non-local, and they proposed that a particular norm of the AGP scales exponentially with the system size and provided strong evidences for this.
On the other hand, for perturbations along exact integrability-preserving directions like varying the anisotropy in the XXZ chain, they found that the AGP norm scales polynomially with the system size.
They also noted that perturbations like the ones here generated by $X_{\text{e}}$ or $X_{\text{bo}}$, while breaking the exact integrability and resulting eventually in the chaotic behavior, nevertheless also have the AGP norm scaling polynomially with system size, and they suggested to exclude such special perturbations when checking for quantum chaos~\footnote{A note of caution here is that while a generator $X_{\text{bo}}$ produces a translationally invariant perturbation in an infinite system and can also be an exact AGP in an appropriately defined finite system with open boundaries, it is not an exact AGP in a finite system with periodic boundary conditions. It would be interesting to study the precise scaling of the AGP norm in this case.}.

Interestingly, in our formalism both the integrability-preserving perturbations and the integrability-breaking perturbations generated by $X_{\text{e}}$, $X_{\text{bo}}$, and $X_{\text{bi}}$ are formally $\ell=2$ weak perturbations, i.e., can be viewed as falling into the same group; likewise, they share a similar polynomial scaling of the AGP norm.
Of course, the latter perturbations will eventually lead to thermalization, and while they do not behave like other ``truly generic'' perturbations, it is also interesting to explore how thermalization happens under such ``less generic'' weak perturbations:
We discuss this in Sec.~\ref{sec:thermtime}, where we argue that the relaxation times have different parametric dependence on the perturbation strength $\lambda$.
Furthermore, our formalism provides recipes to tabulate families of such weak integrability breaking perturbations beforehand, and such tables can then be used to systematically exclude them when needed for studies of more generic thermalization phenomena.
While we do not provide complete tabulations (which is important future work), we showcase many examples; interestingly, already we find that the number of such special perturbations can be significant enough for them to be of practical importance, as discussed earlier.

Our results suggest several intriguing directions for further studies.
One interesting possibility is the study of transport properties of integrable models with weak perturbations \cite{Bastianello_2021,gopalakrishnan2022anomalous}.
Recent developments in the study of transport in integrable systems, in particular within the framework of {\it generalized hydrodynamics}, revealed different regimes, including ballistic, diffusive, and also anomalous superdiffusive \cite{Znidaric2011,Ljubotina2019KPZ,Ilievski2021}.
Small integrability breaking perturbations induce scattering of quasi-particles, affecting the transport properties of the system.
It is an interesting question to understand how weak perturbations compare to ordinary perturbations in this context \cite{Ferreira2020,DeNardis2021}. 
For example, Ref.~\cite{DeNardis2021} showed that the superdiffusive behavior of the Heisenberg model persists up to all numerically accessible times when a small to sizable next-nearest neighbor Heisenberg interaction (a weak integrability breaking perturbation) is included.
However, it is not clear whether the ``weak property'' of the perturbation  plays a role, since similar persistence is observed for other SU$(2)$ preserving perturbations that do not have this property.

One of the possibilities offered by our approach is the ability to systematically compute {\it all} the approximately conserved quantities.
The construction could be applied also to the quasi-local charges, that were first discovered in the XXZ chain~\cite{Ilievski2015}.
It was shown that, in many cases, taking into account these quasi-local charges is crucial to obtain the correct results, for example in the computation of the Drude weight using Mazur bounds, and of the expectation values of the generalized Gibbs ensembles~\cite{Ilievski2015b,Ilievski_2016}.
We expect that the approximately-conserved quasi-local operators that can be computed with our construction are similarly important in determining the properties of weakly-perturbed integrable models.

Another interesting question regards systems with finite size.
We argued that weak perturbations correspond to longer thermalization times in the non-equilibrium dynamics in the thermodynamic limit.
However, the signatures of thermalization are present also in the opposite order of limits~\cite{Bulchandani}:
The transition from an integrable to a thermalizing regime can be captured from the energy spectrum at finite size, for example, by studying the crossover in the level spacing distribution from a Poisson to a Wigner-Dyson statistics.
A recent work \cite{Szasz2021} proposed that a weak perturbation (such as the ones generated by boosted operators) corresponds to a different scaling of the position of the crossover with system size and provided numerical evidence in a perturbed XXZ chain.
The crossover to a Wigner-Dyson statistics was also studied in Ref. \cite{mcloughlin2022chaotic} for the Heisenberg spin chain perturbed with the next-nearest-neighbor interactions, which found that the crossover occurs at larger coupling than for generic perturbations.
It would be interesting to study this for different types of generators $X_{\text{e}}$, $X_{\text{bo}}$, and $X_{\text{bi}}$, for different models and boundary conditions, and also see if there is relation to the AGP norms.

We note that, while we focused on translationally invariant systems, weak integrability breaking perturbations can be also inhomogeneous.
The simplest example is when we turn an extensive local generator $X_\text{e}$ into an inhomogeneous one or even a strictly local operator, which produces an inhomogeneous or a strictly local weak integrability breaking perturbation.
There are also variants starting from boosted generators 
\footnote{
We can turn a boosted generator into a variant producing inhomogeneous or even strictly local perturbations as follows:
Taking $X = \sum_{k \geq j} q_{\beta,k}^{(0)}$, we obtain a perturbation $i[X, Q_2^{(0)}] = -J_{\beta,2;j}$, which is proportional to a local current of the conserved quantity $Q_\beta^{(0)}$ (with convention $H^{(0)} \sim Q_2^{(0)}$).
Such local currents can then be added with independent coefficients, giving weak integrability breaking perturbations of the form $V = \sum_j a_j J_{\beta,2;j}$.},
but so far we have not been able to achieve this starting from bilocal generators.
We leave a systematic study of possible inhomogeneous weak integrability breaking perturbations for future work.

While we focus here on integrable Hamiltonians, some of the most relevant experimental realizations of integrable models are Floquet integrable models realized with circuits of gates \cite{Vanicat2018, Ljubotina2019, Aleiner2021, morvan2022formation}.
It would be interesting to extend our construction to such models: 
While the procedure can be immediately generalized for deformations induced by extensive local operators, it is unclear to us how to generate weak perturbations that preserve the simple gate structure using boosted or bilocal operators. 

We remark that the construction that is the object of our study applies not only to integrable models, but it can also be used to generate weak perturbations of one-dimensional models with a finite number of conserved quantities.
A possible question is whether a similar procedure can generate perturbations that are weak only for a specific subspace, as such perturbations would correspond to models with prominent non-exact quantum many-body scars.
In fact, the tower of scars of the PXP model is an example of such non-exact quantum many-body scars~\cite{Bernien2017, Turner2017, Shiraishi2017, Moudgalya2018a, Serbyn2020review, Papic2021review, Moudgalya2021review, Chandran2022review}, and the mechanism that protects their subspace is still unclear. 
We also note that numerical results on some exact quantum-many body scars~\cite{Lin2019} show that they might be robust to lowest order in the perturbation strength \cite{Lin2020, Surace2021}, suggesting that a notion of {\it weak perturbation} may be formulated also for single eigenstates.
Moreover, a recent experiment has found that some eigenstates of a Floquet integrable model~\cite{Aleiner2021} decay very slowly in the presence of an integrability-breaking perturbation \cite{morvan2022formation}.
It is not yet clear if these states are robust to first order in the perturbation strength.

Finally, while we focused on the case of one-dimensional integrable models, one can think of constructing similar weak perturbations in higher dimensions.
In this case, extensive local and boosted operators can be used as generators~\footnote{Consider, e.g., a model of free spin-1/2 fermions hopping on a Bravais lattice $H^{(0)} = -\sum_{\langle r, r'\rangle} \sum_{s=\uparrow,\downarrow} (a_{r,s}^\dagger a_{r',s} + \mathrm{H.c.})$, which can be viewed as an integrable model in any dimension. 
An operator $X_\text{e} = \sum_r n_{r,\uparrow} n_{r,\downarrow}$ will generate a weak (i.e., near-integrable) perturbation consisting of nearest-neighbor quartic terms $V = i[X_\text{e}, H^{(0)}] = -i\sum_{\langle r, r'\rangle} \sum_{s=\uparrow,\downarrow} (n_{r,-s} - n_{r',-s}) (a_{r,s}^\dagger a_{r',s} - \mathrm{H.c.})$. 
On a square lattice, one can consider perturbations generated by boosted operators along one of the two directions $\hat{x}, \hat{y}$, such as $X_\text{bo} = \sum_{r}\sum_{s=\uparrow,\downarrow} r_x n_{r,s}$.
This generates a perturbation $V = i[X_\text{bo}, H^{(0)}] = i\sum_{r} \sum_{s=\uparrow,\downarrow} (a_{r,s}^\dagger a_{r+\hat{x},s} - \text{H.c.})$.
Note, however, that perturbations obtained from boosted operators of fermion bilinears can only contain fermion bilinears, so the perturbed model is still a free-fermion Hamiltonian.},
but we are not aware of other classes of operators --- analogous to bilocal operators in one dimension --- that would produce physical (i.e., local and extensive) perturbations.



\begin{acknowledgements}
We thank Alvise Bastianello, Anushya Chandran, Sarang Gopalakrishnan, and Cheng-Ju~Lin for insightful discussion.
FMS acknowledges support provided by the U.S.\ Department of Energy Office of Science, Office of Advanced Scientific Computing Research, (DE-SC0020290), by Amazon Web Services, AWS Quantum Program, and by the DOE QuantISED program through the theory  consortium ``Intersections of QIS and Theoretical Particle Physics'' at Fermilab.
OIM acknowledges support by the National Science Foundation through grant DMR-2001186. 
A part of this work was done at the Aspen Center for Physics, which is supported by the National Science Foundation grant PHY-1607611.
\end{acknowledgements}

\bibliography{bib}

\appendix

\section{Details for the Heisenberg chain}

\subsection{Quasi-conserved quantities in the Heisenberg chain perturbed by the second-neighbor interaction}
\label{app:quasiIoMsHeis}
Here we provide some details for the quasi-IoMs in the Heisenberg chain perturbed by the second-neighbor Heisenberg interaction, obtained using generator in Eq.~(\ref{eq:Q1Heis}).

To simplify the notation it is convenient to define a nested vector product operator:
\begin{align}
& F^{i_1 i_2}_j = \vec \sigma_{j+{i_1}} \cdot \vec \sigma_{j+{i_2}} \\
& F^{i_1 i_2 \dots i_n}_j = \{[(\vec \sigma_{j+{i_1}} \times \vec \sigma_{j+{i_2}}) \times \dots ] \times \vec \sigma_{j+{i_{n-1}}}\} \cdot \vec \sigma_{j+{i_n}}.
\end{align}
In the following, we will need the densities for the original conserved charges up to $q_{6,j}^{(0)}$:
\begin{align}
q_{2,j}^{(0)} =& \frac{1}{2} F_j^{01} ~, \qquad q_{3,j}^{(0)} = -\frac{1}{2} F_j^{012} ~, \label{eq:Q20Heis} \\
q_{4,j}^{(0)} =& F_j^{0123} + F_j^{02} - 2F_j^{01} ~, \label{eq:Q40Heis} \\
q_{5,j}^{(0)} =& -3F_j^{01234} - 3F_j^{013} - 3F_j^{023} + 9F_j^{012} ~, \\
q_{6,j}^{(0)} =& 12 F_j^{012345} + 12 F_j^{0124} + 12 F_j^{0134} + 12 F_j^{0234} ~~ \label{eq:Q60Heis} \\
& -48 F_j^{0123} + 12 F_j^{03} - 36 F_j^{02} + 36F_j^{01} ~. \nonumber
\end{align}

While $Q_\alpha^{<2} = Q_\alpha^{(0)} + \lambda Q_\alpha^{(1)}$ are quasi-integrals of motion to order $\ell = 2$ originating from $Q_\alpha^{(0)}$, in analogy to Eqs.~(\ref{eq:Hn}) and (\ref{eq:H1}) for the Hamiltonian we in fact have more freedom in writing down such quasi-IoMs:
\begin{equation}
\tilde{Q}_\alpha^{<2} = Q_\alpha^{(0)} + \lambda Q_\alpha^{(1)} + \lambda \sum_\beta d_{\alpha,\beta}^{(1)} Q_\beta^{(0)} ~,
\label{eq:Qalpha_ell2_gen}
\end{equation}
where $d_{\alpha,\beta}^{(1)}$ can be any fixed real numbers.
Starting with 
\begin{equation}
q_{3, j}^{(1)} = F_j^{01234} + \frac{1}{2} F_j^{013} + \frac{1}{2} F_j^{023} ~,
\end{equation}
the following choice $d_{3,3}^{(1)} = 6$, $d_{3,5}^{(1)} = 1/3$ gives
\begin{equation}
\tilde{q}_{3,j}^{(1)} = q_{3,j}^{(1)} + 6 q_{3,j}^{(0)} + \frac{1}{3} q_{5,j}^{(0)} = -\frac{1}{2}(F_j^{013} + F_j^{023}) ~,
\end{equation}
which matches the quasi-IoM in Eq.~(\ref{eq:tildeQ3}) found in previous works~\cite{Jung2006,Kurlov2022}.

Reference~\cite{Kurlov2022} also calculated several longer-ranged quasi-conserved quantities.
In our approach, we can obtain the next quasi-conserved quantity originating from $Q_4^{(0)}$ using Eq.~(\ref{eq:Q1Heis}), which gives
\begin{align}
q_{4,j}^{(1)} = & -3F_j^{012345} - 2F_j^{0124} - 2F_j^{0134} - 2F_j^{0234} ~~ \\
& + 3F_j^{0123} - F_j^{0213} - 2F_j^{03} - 4F_j^{02} + 18F_j^{01} ~. \nonumber
\end{align}

Note that any linear combination of such $Q_\alpha^{<2}$ is also a valid quasi-IoM, and to match with the next quasi-IoM in Ref.~\cite{Kurlov2022} we need to start with combination $Q_4^{<2} + 4 Q_1^{<2}$ to match their conventions for the original IoMs.
Furthermore, we can add $\lambda$ times any combination of the original IoMs while preserving the $\ell=2$ quasi-conservation.
It is then straightforward to check that the following combination
\begin{align}
& Q_4^{(0)} + 4 Q_2^{(0)} + \lambda \left(Q_4^{(1)} + 4 Q_2^{(1)} \right) + \\
& + \lambda \left[\frac{1}{4} Q_6^{(0)} + (11+a) Q_4^{(0)} + (2+4a) Q_2^{(0)} \right] \nonumber
\end{align}
matches exactly Eq.~(25) in \cite{Kurlov2022}.
They found this quasi-conserved quantity by brute-force search over operators up to fixed range (here range $5$), which is one organizational principle well suited for such search.
This fixes the coefficient of $Q_6^{(0)}$ in the above equation, so as to cancel the range-$6$ term in $Q_4^{(1)}$.
The (linear in $\lambda$) terms proportional to $Q_4^{(0)}$ and $Q_2^{(0)}$ of course can be added with arbitrary coefficients.
Reference~\cite{Kurlov2022} found only one-parameter family given by the above equation with the parameter $a$, probably because their search required the coefficient of $\vec{\sigma}_j \cdot \vec{\sigma}_{j+1}$ in the very final expression to be zero.
While the appearance and count of such free parameters is somewhat mysterious in their formalism, it is not from our perspective where all operators of the form Eq.~(\ref{eq:Qalpha_ell2_gen}) are formally quasi-IoMs to order $\ell=2$ in the sense of Eq.~(\ref{eq:quasicons1}).

Equation~(\ref{eq:Qalpha_ell2_gen}) clearly shows that there can be infinitely many quasi-IoMs associated with each original IoM $Q_\alpha^{(0)}$.
While formally for fixed $\alpha$ they are linearly independent for linearly independent $d_{\alpha,\beta}^{(1)}$, it is not to say that physically they are equally important (and they are not independent when varying $\alpha$).
The perturbative setup keeping $\lambda$ small works well when $\lambda$ multiplies objects that do not get large themselves, and which specific quasi-IoM is best to use can depend on the context (e.g., Ref.~\cite{Kurlov2022} used conditions of fixed range and minimization of the Frobenius norm of the commutator with the perturbed Hamiltonian, which is reasonable for certain types of quenches~\cite{Kim2015}).

\subsection{Details of constructing $\ell=3$ nearly-integrable perturbation to the Heisenberg chain}
\label{subapp:ell3Heis}
Here we provide some details for constructing particularly simple higher-order nearly-integrable perturbation to the Heisenberg chain.
To evaluate Eq.~(\ref{eq:HeisQ22}), we first calculate
\begin{align}
& -i[{\mathcal B}[Q_3^{(0)}], Q_2^{(1)}] - i[{\mathcal B}[Q_3^{(1)}], Q_2^{(0)}] \\
& = \sum_j \Big( \frac{5}{2} F^{012345}_j + \frac{3}{2} F_j^{0124} + \frac{3}{2} F_j^{0134} + \frac{5}{2} F_j^{0123} \nonumber \\
& \qquad\quad~ + F_j^{0213} + F_j^{03} + 3F_j^{02} - 4F_j^{01} \Big) ~. \nonumber
\end{align}
Note that $i[{\mathcal B}[Q_3^{(0)}], Q_2^{(1)}]$ and $i[{\mathcal B}[Q_3^{(1)}], Q_2^{(0)}]$ are in general {\it not} extensive local operators, since $[Q_3^{(0)}, Q_2^{(1)}] \neq 0$ and $[Q_3^{(1)}, Q_2^{(0)}] \neq 0$.
However, their sum is properly extensive as exhibited above, in agreement with the expectation that $Q_2(\lambda)$ produced by the flow Eq.~(\ref{eq:deform}) with the generators Eq.~(\ref{eq:Xbo}) is extensive order-by-order in $\lambda$.

Next, we choose the coefficients $f_\beta^{(1)}$ to make Eq.~(\ref{eq:HeisQ22}) more simple.
Taking
\begin{equation}
f_\beta^{(1)} = 10 \delta_{\beta,3} + \frac{1}{2} \delta_{\beta,5}    
\end{equation}
and using the already calculated $-i[{\mathcal B}[Q_3^{(0)}], Q_2^{(0)}] = Q_{2}^{(1)}$ from Eq.~(\ref{eq:q21Heis}) and
\begin{align}
& -i[{\mathcal B}[Q_5^{(0)}], Q_2^{(0)}] =
\sum_j \Big( -3F_j^{012345} - 3F_j^{0124} \\
&\qquad\qquad -3F_j^{0134} + 3F_j^{0123} - 6F_j^{02} + 18F_j^{01} \Big) \nonumber
\end{align}
we obtain for Eq.~(\ref{eq:HeisQ22})
\begin{equation}
Q_2^{(2)} = \sum_j \Big( F_j^{012345} - F_j^{0123} + F_j^{0213} + F_j^{03} - 5F_j^{01} \Big).
\end{equation}

Finally, we choose coefficients $c_\alpha^{(2)}$ to make $H^{(2)}$ in Eq.~(\ref{eq:H22_Heis_ell3}) as simple as possible:
\begin{equation}
c_\alpha^{(2)} = 36 \delta_{\alpha,2} + 20 \delta_{\alpha,4} + \frac{1}{3} \delta_{\alpha,6} ~.
\end{equation}
Using expressions for $Q_{2}^{(0)}$, $Q_{4}^{(0)}$, and $Q_6^{(0)}$ [Eqs.~(\ref{eq:Q20Heis}), (\ref{eq:Q40Heis}), and (\ref{eq:Q60Heis})], we 
find
\begin{equation}
H^{(2)} = \sum_j 2F_j^{03} = \sum_j 2 \vec{\sigma}_j \cdot \vec{\sigma}_{j+3} ~,
\end{equation}
which contains only the simple third-neighbor Heisenberg interaction.
Plugging this $H^{(2)}$ into Eq.~(\ref{eq:Hl3gen}) gives the claimed $H^{<3}(\lambda)$ in Eq.~(\ref{eq:Hl3Heis}).

\section{Some details for the XYZ and XXZ chains}
\label{app:XYZ}
We report here the expressions for the deformations $Q_2^{(1)}$ and $Q_3^{(1)}$ generated by $\mathcal B[Q_3^{(0)}]$ in the XYZ chain in Sec.~\ref{subsec:XYZ}:
We start with $Q_2^{(1)} = -i[{\mathcal B}[Q_3^{(0)}], Q_2^{(0)}]$, which is a nearly-integrable perturbation to the XYZ chain; the density reads:
\begin{align} 
q_{2,j}^{(1)} = & \frac{1}{2} \sum_{(\alpha,\beta,\gamma) \in \Pi} t_\alpha^2 t_\gamma \sigma_j^\alpha \sigma_{j+1}^\beta \sigma_{j+2}^\beta \sigma_{j+3}^\alpha \nonumber \\
& - \frac{1}{2}\, t_x t_y t_z \! \sum_{(\alpha,\beta,\gamma) \in \Pi}  \sigma_j^\alpha \sigma_{j+1}^\beta \sigma_{j+2}^\alpha \sigma_{j+3}^\beta \nonumber \\ 
&- \frac{1}{2}\sum_{\alpha \in \{x,y,z\}}t_\alpha (\vec{t}^{\;2}-t_\alpha^2)\sigma_j^\alpha \sigma_{j+1}^\alpha ~, \nonumber
\end{align}
where $\Pi$ in $\sum_{(\alpha,\beta,\gamma) \in \Pi}$ is the set of permutations of $(x,y,z)$ (six terms in the sum).
This nearly-integrable perturbation specialized to the XXZ chain ($t_x=t_y=1$, $t_z=\Delta$) was considered in Ref.~\cite{Szasz2021}, see Eq.~(6) there 
\footnote{There is a small typo in their expression for $j_{3,l}$ in what should be $xzzx$ term in the third line, and our $q_{2,j}^{(1)}$ agrees with the $j_{3,l}$ if their square bracket closes after the third line rather than the fourth one, although there may also be a difference in some conventions}.
As we discuss in the main text, by combining with $Q_4^{(0)}$ we can obtain a simpler nearly-integrable perturbation, which in turn can be traded for a yet simpler perturbation by removing the nearest-neighbor $\sigma_j^\alpha \sigma_{j+1}^\alpha$ terms, obtaining only the second-neighbor Heisenberg interaction $\vec{\sigma}_j \cdot \vec{\sigma}_{j+2}$.

Next, we calculate $Q_3^{(1)} = -i[{\mathcal B}[Q_3^{(0)}], Q_3^{(0)}]$, which gives appropriate correction to $Q_3^{(0)}$ to obtain quas-IoM for the above nearly-integrable perturbations.
The density reads:
\begin{align}
& q_{3,j}^{(1)} = -\sum_{\alpha,\beta,\gamma} \epsilon_{\alpha\beta\gamma} t_\alpha^2 t_\gamma^2 \sigma_j^\alpha \sigma_{j+1}^\beta \sigma_{j+2}^\beta \sigma_{j+3}^\beta \sigma_{j+4}^\gamma \\ 
& + t_x t_y t_z \sum_{\alpha,\beta,\gamma} \epsilon_{\alpha\beta\gamma} t_\alpha \sigma_j^\alpha \sigma_{j+1}^\beta \sigma_{j+2}^\alpha \sigma_{j+3}^\gamma \sigma_{j+4}^\alpha \nonumber \\
& + t_x t_y t_z \sum_{\alpha,\beta,\gamma} \epsilon_{\alpha\beta\gamma} t_\alpha \sigma_j^\alpha \sigma_{j+1}^\beta \sigma_{j+2}^\beta \sigma_{j+3}^\gamma \sigma_{j+4}^\beta \nonumber \\
& - t_x t_y t_z \sum_{\alpha,\beta,\gamma} \epsilon_{\alpha\beta\gamma} t_\gamma \sigma_j^\alpha \sigma_{j+1}^\beta \sigma_{j+2}^\alpha \sigma_{j+3}^\alpha \sigma_{j+4}^\gamma \nonumber \\
& + \frac{1}{2} t_x t_y t_z \sum_{\alpha,\beta,\gamma} \epsilon_{\alpha\beta\gamma} (t_\alpha \sigma_j^\alpha \sigma_{j+1}^\beta \sigma_{j+3}^\gamma + t_\gamma \sigma_j^\alpha \sigma_{j+2}^\beta \sigma_{j+3}^\gamma) ~. \nonumber
\end{align}
As we show in the main text, an appropriate combination with $Q_5^{(0)}$ eliminates all the range-4 terms giving a simpler quasi-IoM expression.

\section{Simplifying structures for weak perturbations generated by $\hat{X} \sim {\mathcal B}[\hat{Q}_3]$ in integrable models satisfying Reshetikhin criterion}
\label{app:detailsBQ3}

In this appendix, we provide an explanation for the cancellation of longer-range terms in the nearly-integrable perturbations in the Heisenberg and XYZ chains constructed using generator $X \sim {\mathcal B}[Q_3^{(0)}]$ in Secs.~\ref{subsec:Heisenberg} and \ref{subsec:XYZ}, showing that this structure applies for more general integrable models satisfying certain conditions.
This also gives an alternative derivation of the corresponding results for the Heisenberg and XYZ chains.

Consider a 1d Hamiltonian with two-site terms
\begin{equation}
H_0 = \sum_j h_{j,j+1} ~.
\end{equation}
Here and below, to make appearing structures more clear, we use notation showing explicitly all sites involved in each term; we will connect with the notation used in the main text when appropriate.
We have
\begin{align}
[h_{j,j+1}, H_0] &= [h_{j,j+1}, h_{j+1,j+2}] - [h_{j-1,j}, h_{j,j+1}] \\
&\equiv g_{j,j+1,j+2} - g_{j-1,j,j+1} ~.
\end{align}
Hence, the commutator of the boosted Hamiltonian with the Hamiltonian is an extensive local operator:
\begin{equation}
\Big[ \sum_j j h_{j,j+1}, H_0 \Big] = -\sum_j g_{j,j+1,j+2} ~.
\end{equation}
The underlying reason for this is the energy conservation, and the R.H.S.\ is (proportional to) a sum of local energy currents.
The above is valid for any Hamiltonian.

From now on we will consider integrable Hamiltonians whose IoMs can be generated using the boosted Hamiltonian as $Q_{n+1} \sim [{\mathcal B}[H_0], Q_n]$, $n \geq 2$, with $Q_2 \sim H_0$, as in the Heisenberg and XYZ spin chains in the main text.
Following Ref.~\cite{Grabowski1995IntTest}, let us consider when thus defined $Q_3$ can commute with $H_0$.
We have
\begin{align}
& [h_{j,j+1}, Q_3] \sim \Big[ h_{j,j+1}, \sum_{j'} g_{j',j'+1,j'+2} \Big] = \label{eq:comm_h_Q3} \\
& = [h_{j,j+1}, g_{j+1,j+2,j+3}] - [h_{j-2,j-1}, g_{j-1,j,j+1}] \nonumber \\
& + [h_{j,j+1}, g_{j,j+1,j+2}] + [h_{j,j+1}, g_{j-1,j,j+1}] ~, \nonumber
\end{align}
where we used the expression for $g$'s in terms of $h$'s to rewrite $[h_{j,j+1}, g_{j-2,j-1,j}] = - [h_{j-2,j-1}, g_{j-1,j,j+1}]$ (since $h_{j-2,j-1}$ and $h_{j,j+2}$ commute).
To evaluate $[H_0, Q_3]$ we need to sum Eq.~(\ref{eq:comm_h_Q3}) over $j$.
The first two terms have a ``telescoping structure,'' hence their contributions will cancel upon the summation.
No such structure is present for the last two terms for general Hamiltonians; however, we will have the telescoping structure if so-called Reshetikhin condition~\cite{Grabowski1995IntTest} is satisfied:
There exist two-site operators $R_{j,j+1}$ such that
\begin{equation}
[h_{j,j+1} + h_{j+1,j+2}, g_{j,j+1,j+2}] = R_{j,j+1} - R_{j+1,j+2} ~. \label{eq:Reshetikhin}
\end{equation}
This condition is satisfied for the Heisenberg and XYZ chains, and in general it implies $[H_0, Q_3] = 0$.

If the Reshetikhin condition is satisfied, then it is also clear that $Q_4$ as defined above is an extensive local operator.
A straightforward calculation using Eqs.~(\ref{eq:comm_h_Q3}) and (\ref{eq:Reshetikhin}) gives
\begin{align}
& \Big[ \sum_j j h_{j,j+1}, \sum_{j'} g_{j',j'+1,j'+2} \Big] = \label{eq:propQ4} \\
& = \sum_j \Big( -2 [h_{j,j+1}, g_{j+1,j+2,j+3}] + \nonumber \\
& ~~~~~~~~~~~ + [h_{j+1,j+2}, g_{j,j+1,j+2}] + R_{j,j+1} \Big) ~. \nonumber
\end{align}

In the main text we considered perturbed models generated using $X \sim {\mathcal B}[Q_3]$, which led to special perturbations $\sim [{\mathcal B}[Q_3], H_0]$.
Reusing some calculations behind Eq.~(\ref{eq:comm_h_Q3}), we have
\begin{align}
& \Big[ h_{j,j+1}, \sum_{j'} j' g_{j',j'+1,j'+2} \Big] = \\
& = (j+1) [h_{j,j+1}, g_{j+1,j+2,j+3}] \nonumber \\
& ~~~ - (j-2) [h_{j-2,j-1}, g_{j-1,j,j+1}] \nonumber \\
& ~~~ + j [h_{j,j+1}, g_{j,j+1,j+2}] + (j-1) [h_{j,j+1}, g_{j-1,j,j+1}] ~. \nonumber
\end{align}
Summation over $j$ using the telescoping structure in the first two terms and the Reshetikhin condition for the second two terms then gives
\begin{align}
& \Big[ \sum_j h_{j,j+1}, \sum_{j'} j' g_{j',j'+1,j'+2} \Big] = \label{eq:propCurrQ3} \\
& = \sum_j \Big([h_{j,j+1}, g_{j+1,j+2,j+3}] + R_{j,j+1} \Big) ~. \nonumber
\end{align}
We see that in this total current associated with the conserved quantity $Q_3$, the range-4 terms have exactly the same structure as in $Q_4$ in Eq.~(\ref{eq:propQ4}).
Hence we can obtain a simpler nearly-integrable perturbation by combining Eqs.~(\ref{eq:propCurrQ3}) and (\ref{eq:propQ4}):
\begin{align}
& \Big[ \sum_j h_{j,j+1}, \sum_{j'} j' g_{j',j'+1,j'+2} \Big] + \nonumber \\
& + \frac{1}{2} \Big[ \sum_j j h_{j,j+1}, \sum_{j'} g_{j',j'+1,j'+2} \Big] = \nonumber \\
& = \sum_j \Big(\frac{1}{2} [h_{j+1,j+2}, g_{j,j+1,j+2}] + \frac{3}{2} R_{j,j+1} \Big) ~.
\label{eq:app_simpler_combo}
\end{align}

Connecting with the notation in the main text used in the Heisenberg and XYZ cases, Secs.~\ref{subsec:Heisenberg} and \ref{subsec:XYZ}, we have:
\begin{align}
Q_2^{(0)} &\equiv \frac{1}{2} H_0 = \frac{1}{2} \sum_j h_{j,j+1} ~, \\
Q_3^{(0)} &\equiv i \Big[\sum_j j q_{2,j}^{(0)}, Q_2^{(0)} \Big] = -\frac{i}{4} \sum_j g_{j,j+1,j+2} ~, \\
Q_4^{(0)} &\equiv i \Big[\sum_j j q_{2,j}^{(0)}, Q_3^{(0)} \Big] \\
&= \frac{1}{8} \Big[ \sum_j j h_{j,j+1}, \sum_{j'} g_{j',j'+1,j'+2} \Big] ~, \nonumber \\
Q_2^{(1)} &\equiv -i\Big[\sum_{j'} j' q_{3,j'}^{(0)}, Q_2^{(0)} \Big] \\
&= -\frac{1}{8} \Big[ \sum_{j'} j' g_{j',j'+1,j'+2}, \sum_j h_{j,j+1} \Big] ~. \nonumber 
\end{align}
In the preceding discussions and expressions, everything is still general other than the convention $Q_2^{(0)} \equiv H_0/2$,  i.e., applicable for any integrable model satisfying the Reshetikhin condition.
The specific combination of $Q_2^{(1)}$ and $Q_4^{(0)}$ used to obtain more simple nearly-integrable perturbations in the Heisenberg and XYZ models in the main text, Eqs.~(\ref{eq:H1Heis2ndnb}) and (\ref{eq:XYZH1}) respectively, in fact works for all such integrable models, since it is precisely the combination in Eq.~(\ref{eq:app_simpler_combo}) that cancels the four-site terms:
\begin{equation}
2 Q_2^{(1)} + Q_4^{(0)} = \frac{1}{4} \sum_j \Big( \frac{1}{2} [h_{j+1,j+2}, g_{j,j+1,j+2}] + \frac{3}{2} R_{j,j+1} \Big) ~.
\label{eq:app_Q21_Q40_combo}
\end{equation}

We can use these formulas to obtain alternative derivations of the final results in the Heisenberg and XYZ chain, which we provide here for easy reference.
Specializing to the case of the Heisenberg chain, $h_{j,j+1} = \vec{\sigma}_j \cdot \vec{\sigma}_{j+1}$, we have
\begin{align}
& g_{j,j+1,j+2} = -2i (\vec{\sigma}_j \times \vec{\sigma}_{j+1}) \cdot \vec{\sigma}_{j+2} ~, \\
& [h_{j+1,j+2}, g_{j,j+1,j+2}] = 8 (\vec{\sigma}_j \cdot \vec{\sigma}_{j+2} - \vec{\sigma}_j \cdot \vec{\sigma}_{j+1}) ~, \\
& R_{j,j+1} = -8 \vec{\sigma}_j \cdot \vec{\sigma}_{j+1} ~.  
\end{align}
Plugging this into Eq.~(\ref{eq:app_Q21_Q40_combo}) reproduces the result in the main text, Eq.~(\ref{eq:H1Heis2ndnb}), where in addition we removed the nearest-neighbor Heisenberg term by taking appropriate combination with $Q_2^{(0)}$ (this additional simplification is special for the Heisenberg model).

In the XYZ chain case, $h_{j,j+1} = \sum_\alpha t_\alpha \sigma_j^\alpha \sigma_{j+1}^\alpha$, we have:
\begin{align}
& g_{j,j+1,j+2} = -2i 
\sum_{\alpha\beta\gamma} \epsilon_{\alpha\beta\gamma} t_\alpha t_\gamma \sigma_j^\alpha \sigma_{j+1}^\beta \sigma_{j+2}^\gamma ~, \\
& [h_{j+1,j+2}, g_{j,j+1,j+2}] = 8 t_x t_y t_z \, \vec{\sigma}_j \cdot \vec{\sigma}_{j+2} - \\
& ~~~~~ - 4 \sum_\alpha t_\alpha (\vec{t}^{\;2} - t_\alpha^2) \sigma_j^\alpha \sigma_{j+1}^\alpha  ~, \\
& R_{j,j+1} = -4 \sum_\alpha t_\alpha (\vec{t}^{\;2} - t_\alpha^2) \sigma_j^\alpha \sigma_{j+1}^\alpha ~.
\end{align}
Plugging this into Eq.~(\ref{eq:app_Q21_Q40_combo}) reproduces the XYZ chain result in the main text, Eq.~(\ref{eq:XYZH1}).

\section{Hubbard model}
\label{app:Hubbard}
In this Appendix, we collect some details behind the Hubbard model results in Sec.~\ref{subsec:Hubbard}.
First, we list the IoM densities $q_{2,j}^{(0)}$ and $q_{3,j}^{(0)}$ that are respectively even and odd under inversion in bond center between sites $j$ and $j+1$:
\begin{eqnarray}
q_{2,j}^{(0)} &\equiv& -2\sum_{s=\uparrow,\downarrow} (a_{j,s}^\dagger a_{j+1,s} + \text{H.c.}) \label{eq:q2Hubb} \\
&+& 2U \left(n_{j,\uparrow}-\frac{1}{2}\right) \left(n_{j,\downarrow}-\frac{1}{2}\right) \nonumber \\
&+& 2U \left(n_{j+1,\uparrow}-\frac{1}{2}\right) \left(n_{j+1,\downarrow}-\frac{1}{2}\right) ~; \nonumber
\end{eqnarray}
\begin{eqnarray}
q_{3,j}^{(0)} &\equiv& -i\sum_{s} (a_{j,s}^\dagger a_{j+2,s} - \text{H.c.}) \label{eq:q3Hubb} \\
&-& i\sum_{s} (a_{j-1,s}^\dagger a_{j+1,s} - \text{H.c.}) \nonumber \\
&+& 4 i U \sum_{s} (a_{j,s}^\dagger a_{j+1,s} - \text{H.c.})(n_{j,-s} + n_{j+1,-s} - 1) ~. \nonumber
\end{eqnarray}
We also list the next IoM of the Hubbard model from Ref.~\cite{Grabowski1995}:
\begin{eqnarray}
Q_4^{(0)}&=&\sum_{j,s} \left[-2(a_{j,s}^\dagger a_{j+3,s}+\text{H.c.})\right. \label{eq:HubbQ4} \\
&+& 4U(a_{j,s}^\dagger a_{j+2,s}+\text{H.c.})\times \nonumber\\
&& \qquad\times\left(n_{j,-s}+n_{j+1,-s}+n_{j+2,-s}-\frac{3}{2}\right) \nonumber\\
&+& 2U(a_{j,s}^\dagger a_{j+1,s}-\text{H.c.})(a_{j,-s}^\dagger a_{j+1,-s}-\text{H.c.}) \nonumber\\
&+& 4U(a_{j,s}^\dagger a_{j+1,s}-\text{H.c.})(a_{j+1,-s}^\dagger a_{j+2,-s}-\text{H.c.}) \nonumber\\
&-& 4U\left(n_{j,s}-\frac{1}{2}\right)\left(n_{j+1,-s}-\frac{1}{2}\right) \nonumber\\
&-& 2U\left(n_{j,s}-\frac{1}{2}\right)\left(n_{j,-s}-\frac{1}{2}\right) \nonumber\\
&-& 4U^2(a_{j,s}^\dagger a_{j+1,s}+\text{H.c.})\times \nonumber\\
&& \qquad \times(2n_{j,-s} n_{j+1,-s}-n_{j,-s}-n_{j+1,-s} + 1)\Big].\nonumber
\end{eqnarray}

This IoM can be used to eliminate the third-neighbor hopping term from the weak integrability breaking perturbation in the main text, Eq.~(\ref{eq:HubbJ32tot}), but at the expense of introducing additional terms, including the six-fermion terms in the last line.

We can eliminate the six-fermion terms by considering weak integrability breaking perturbation generated by the following extensive local term:
\begin{equation}
X = 4i\sum_{j,s}(a_{j,s}^\dagger a_{j+1,s} - \text{H.c.}) (n_{j,-s} - n_{j+1,-s}) ~.
\end{equation}
This $X$ is manifestly invariant under the inversion but is odd under the time reversal, hence the corresponding $V = i[X, Q_{2}^{(0)}]$ is invariant under both these symmetries.
One can also verify that $X$ has both the spin SU(2) and pseudo-spin SU(2) symmetries, hence $V$ has these symmetries too.
Evaluating this $V$ and constructing an appropriate combination with $Q_{4}^{(0)}$, we can then
eliminate the six-fermion terms, and an appropriate combination with $J_{3,2;\text{tot}}^{(0)}$, Eq.~(\ref{eq:HubbJ32tot}), then gives a weak integrability breaking perturbation that is of range 3 and has only four-fermion terms:
\begin{align}
& J_{3,2;\text{tot}}^{(0)} - 2Q_4^{(0)} - \frac{1}{2} U i[X, Q_2^{(0)}] = \label{eq:Hubbard_simplest_pert} \\ 
&= \sum_{j,s}(4+8U^2)(a_{j,s}^\dagger a_{j+1,s}+\text{H.c.}) \nonumber \\
&~ +8 U\sum_{j}\left(n_{j,\uparrow}-\frac{1}{2}\right) \left(n_{j,\downarrow}-\frac{1}{2}\right) \nonumber \\ 
&~ -8U \sum_{j,s} (a_{j,s}^\dagger a_{j+1,s} - \text{H.c.})(a_{j+1,-s}^\dagger a_{j+2,-s} - \text{H.c.}) \nonumber \\
&~ -4U \sum_{j,s} (a_{j,s}^\dagger a_{j+2,s}+\text{H.c.})(2n_{j+1,-s}-1) \nonumber \\
&~ +4U \sum_{j,s} (a_{j,s}^\dagger a_{j+1,s}-\text{H.c.}) (a_{j,-s}^\dagger a_{j+1,-s}-\text{H.c.}) \nonumber \\
&~ -8U \sum_{j,s} \left(n_{j,s}-\frac{1}{2}\right) \left(n_{j+1,-s}-\frac{1}{2}\right). \nonumber
\end{align}
Note that we can drop the nearest-neighbor hopping and the on-site Hubbard terms and still maintain the weak integrability breaking property (the quasi-IoMs would need to be adjusted when this is done).
The final expression is then a relatively simple one containing particular combinations of current-current interactions for opposite spin species  on neighboring links, as well as correlated second-neighbor hopping terms.
The nearest-neighbor terms in the last two lines can be combined together using the identity
\begin{align*}
& 2(\vec{S}_j \cdot \vec{S}_{k} - \vec{T}_j \cdot \vec{T}_{k}) = (a_{j\uparrow}^\dagger a_{k\uparrow} - \text{H.c.}) (a_{j\downarrow}^\dagger a_{k\downarrow} - \text{H.c.}) - \\
& - \left(n_{j\uparrow} - \frac{1}{2} \right) \left(n_{k\downarrow} - \frac{1}{2} \right)
- \left(n_{j\downarrow} - \frac{1}{2} \right) \left(n_{k\uparrow} - \frac{1}{2} \right) ~,
\end{align*}
where $\vec{S}_j$ is the on-site spin operator [$S_j^+ \equiv a_{j\uparrow}^\dagger a_{j\downarrow}, S_j^z \equiv \frac{1}{2}(n_{j\uparrow} - n_{j\downarrow})$], while $\vec{T}_j$ is the on-site pseudo-spin appropriate for real hopping on the bipartite lattice [$T_j^+ \equiv (-1)^j c_{j\uparrow}^\dagger c_{j\downarrow}^\dagger, T_j^z \equiv \frac{1}{2}(n_{j\uparrow} + n_{j\downarrow} - 1)$], and the sites $j$ and $k$ are on the opposite sublattices.
This form is manifestly spin SU(2) and pseudo-spin SU(2) invariant.

\section{Quasi-IoMs for nearly-integrable perturbations of free spinless fermions}
\label{app:freefer}
Here we provide the expressions of the deformations of the quantity $Q_3'$ generated by the bilocal operators considered in Sec.~\ref{subsec:freeferms}:
\begin{align}
(a)\!:\!& \quad i[[Q_2'^{(0)}|Q_1'^{(0)}], Q_3'^{(0)}] = -2\sum_j q_{4,j}''^{(0)} \!\cdot\! (n_{j+1}+n_{j+2}) \nonumber\\
& \hspace{2.5cm} -2\sum_j q_{2,j}''^{(0)} \!\cdot\! (n_{j-1}+n_{j+2}),\\
(b)\!:\!& \quad i[[Q_2''^{(0)}|Q_1'^{(0)}], Q_3'^{(0)}] = 2\sum_j q'^{(0)}_{4,j} \!\cdot\! (n_{j+1}+n_{j+2}) \nonumber\\
& \hspace{2.5cm} -2\sum_j q_{2,j}'^{(0)} \!\cdot\! (n_{j-1}+n_{j+2}),\\
(c)\!:\!& \quad i[[Q_2'^{(0)}| Q_2''^{(0)}], Q_3'^{(0)}] \nonumber\\
&\hspace{1.8cm} = -2\sum_j q_{3,j}'^{(0)} \!\cdot\! (n_{j-1} - 2n_{j+1} + n_{j+3}) \nonumber\\
& \hspace{1.8cm} +2\sum_j (q_{2,j}'^{(0)} q_{2,j+2}'^{(0)} - q_{2,j}''^{(0)} q_{2,j+2}''^{(0)}).
\end{align}
These are then corrections to $Q_3'^{(0)}$ to produce quasi-IoMs for the corresponding quasi-integrable perturbations in the main text, and one can in principle calculate corrections to all $Q_\beta'^{(0)}$ and $Q_\beta''^{(0)}$.

\section{Thermalization time}
\label{sec:thermtime}

\subsection{Rigorous bound on the thermalization time for nearly-conserved local observables}
Suppose $H$ and $M$ are extensive local and translationally invariant operators,
\begin{equation}
H = \sum_j h_j ~, \quad M = \sum_j m_j ~,
\end{equation}
with local terms $h_j$ and $m_j$ such that the $(j+1)$-th terms are translations of the $j$-th terms.
(Here and below we assume the translational invariance only to simplify extraction of local parts of the observables, but this is not fundamental and can be relaxed under physically reasonable settings working with extensive local operators.)
For applications in the settings in the main text, $H$ is the full perturbed Hamiltonian with the perturbation characterized by the smallness parameter $\lambda$, while $M$ is an approximately conserved quantity that commutes with $H$ up to $O(\lambda^\ell)$ corrections, i.e.,
\begin{equation}
i[H, M] = \lambda^{\ell} R = \lambda^{\ell} \sum_j r_j ~,
\label{eq:HMcommutator}
\end{equation}
where the local terms $r_j$ may have additional $\lambda$ dependence in them but have a bounded operator norm, $\| r_j \| \leq c_r$ with $c_r$ an $O(1)$ number independent of $\lambda$, for some finite range of $\lambda$ near $0$.
Then, for generic quench experiments the thermalization time is rigorously upper-bounded by $O(1/\lambda^{\ell})$.

Indeed, consider quenching from some simple translationally invariant initial state 
$\ket{\Psi_\text{ini}}$, e.g., a product state.
At time $t$ we have
\begin{eqnarray}
&& \bra{\Psi(t)} m_j \ket{\Psi(t)} - \bra{\Psi_\text{ini}} m_j \ket{\Psi_\text{ini}} \nonumber \\ 
&& = \frac{1}{L} \left( \bra{\Psi_\text{ini}} e^{i H t} M e^{-i H t} \ket{\Psi_\text{ini}} - \bra{\Psi_\text{ini}} M \ket{\Psi_\text{ini}} \right) \nonumber \\
&& = \frac{1}{L} \bra{\Psi_\text{ini}} \int_0^t dt'\, e^{i H t'} i [H, M] e^{-i H t'} \ket{\Psi_\text{ini}} \nonumber \\
&& = \lambda^{\ell} \int_0^t dt'\, \bra{\Psi_\text{ini}} e^{i H t'} r_j e^{-i H t'} \ket{\Psi_\text{ini}} ~,
\label{eq:evol_localm}
\end{eqnarray}
where in the intermediate steps $L$ is the system size and we have used the translational invariance of $H$, $M$, $R$, and $\ket{\Psi_\text{ini}}$.
We have $| \bra{\Psi_\text{ini}} e^{i H t'} r_j e^{-i H t'} \ket{\Psi_\text{ini}} | \leq \| r_j \| \leq c_r$, and hence
\begin{eqnarray}
| \bra{\Psi(t)} m_j \ket{\Psi(t)} - \bra{\Psi_\text{ini}} m_j \ket{\Psi_\text{ini}} | \leq 
\lambda^{\ell} c_r t ~.
\end{eqnarray}
Since we expect that the ``thermalized'' value of the local observable at long time, $\lim_{t \to \infty} \bra{\Psi(t)} m_j \ket{\Psi(t)}$, differs from the initial value, $\bra{\Psi_\text{ini}} m_j \ket{\Psi_\text{ini}}$, by a non-zero $O(1)$ amount $\Delta m_{\text{th}}$, we conclude that it will take at least time
\begin{equation}
t \geq \frac{|\Delta m_{\text{th}}|}{c_r \lambda^{\ell}}
\end{equation}
to thermalize, which is parametrically large for small $\lambda$.
Note that this is a completely rigorous lower bound on the thermalization time valid for any $L$ (hence also in the thermodynamic limit $L \to \infty$) that makes no assumptions about the thermalization physics of $H$ and $M$ other than their extensive local character and the near-commutation Eq.~(\ref{eq:HMcommutator}).
Note also that no assumption is made about the initial state $\ket{\Psi_\text{ini}}$ with respect to $H$ and $M$; the bound on the rate of change of the observable $m_j$ is always valid, and the ``thermalization'' assumption is only made to convert the rate to a finite time assuming an $O(1)$ change in the observable under the dynamics.

\subsection{Vanishing of the rate of change of $M$ at $t=0$ for eigenstates of the unperturbed $H^{(0)}$ and $M^{(0)}$}
In the above, the initial state $\ket{\Psi_\text{ini}}$ can be arbitrary, and in such a general setting we are not able to make any further arguments about the thermalization time.
We will now specialize to the case $\ell = 2$, with 
\begin{align}
& H = H^{(0)} + \lambda H^{(1)} ~, \quad M = M^{(0)} + \lambda M^{(1)} ~, \\
& [H^{(0)}, M^{(0)}] = 0 ~,
\end{align}
and consider $\ket{\Psi_\text{ini}} = \ket{\Psi_0}$ which is a simultaneous eigenstate of the unperturbed $H^{(0)}$ and $M^{(0)}$ \footnote{The results below easily generalize to mixed initial states $\rho_{\text{ini}} = \rho_0$ such that $[\rho_0, H^{(0)}] = [\rho_0, M^{(0)}] = 0$, e.g., equilibrium ensembles generated by $H^{(0)}$ and $M^{(0)}$.}
\footnote{Connection to standard notation for the unpertubed Hamiltonian and the perturbation is $H^{(0)} \leftrightarrow H_0$, $H^{(1)} \leftrightarrow V$; we use the current notation to emphasize similarity between $H$ and $M$ in manipulations below and also to connect with notation for quasi-conserved quantities in the main text: $Q^{(0)} \leftrightarrow M^{(0)}$, $Q^{(1)} \leftrightarrow M^{(1)}$.}. 
The commutation condition Eq.~(\ref{eq:HMcommutator}) gives:
\begin{equation}
[H^{(0)}, M^{(1)}] = [M^{(0)}, H^{(1)}] ~, \quad
R \equiv i[H^{(1)}, M^{(1)}] ~.
\label{eq:H0M1_M0H1}
\end{equation}

The rate of change of the quasi-conserved quantity $M$ has $O(\lambda^2)$ smallness prefactor:
\begin{equation}
\frac{d}{dt} \bra{\Psi(t)} M \ket{\Psi(t)} = \lambda^2 \bra{\Psi(t)} R \ket{\Psi(t)} ~.
\label{eq:MRate}
\end{equation}
We will now show that for initial states that are eigenstates of $H^{(0)}$ and $M^{(0)}$, under some additional natural assumptions,
\begin{equation}\bra{\Psi_0} R \ket{\Psi_0} = 0 ~,
\label{eq:Rval0}
\end{equation}
and hence the rate of change of $M$ vanishes for $t=0$.

Consider an orthonormal basis $\{ \ket{\phi_k^{(0)}} \}$ of states that simultaneously diagonalize $H^{(0)}$ and $M^{(0)}$ with eigenvalues $\{ \epsilon_k^{(0)} \}$ and $\{ \mu_k^{(0)} \}$ respectively.
Denote the corresponding matrix elements of $H^{(1)}$ and $M^{(1)}$ as $H^{(1)}_{kn} \equiv \bra{\phi_k^{(0)}} H^{(1)} \ket{\phi_n^{(0)}}$ and $M^{(1)}_{kn} \equiv \bra{\phi_k^{(0)}} M^{(1)} \ket{\phi_n^{(0)}}$.
Condition Eq.~(\ref{eq:H0M1_M0H1}) implies
\begin{equation}
(\epsilon_k^{(0)} - \epsilon_n^{(0)}) M^{(1)}_{kn} = (\mu_k^{(0)} - \mu_n^{(0)}) H^{(1)}_{kn}
\end{equation}
for all $k$ and $n$.
Hence we have
\begin{align}
& \bra{\phi_k^{(0)}} R \ket{\phi_k^{(0)}} = i \sum_n \left( H^{(1)}_{kn} M^{(1)}_{nk} - M^{(1)}_{kn} H^{(1)}_{nk} \right) \nonumber \\
& = i \!\! \sum_{n,~ \epsilon_n^{(0)} = \epsilon_k^{(0)},~ \mu_n^{(0)} = \mu_k^{(0)}} \!\! \left( H^{(1)}_{kn} M^{(1)}_{nk} - M^{(1)}_{kn} H^{(1)}_{nk} \right) ~.
\label{eq:Rmatrelem}
\end{align}
In particular we see that if either $\epsilon_k^{(0)}$ or $\mu_k^{(0)}$ is non-degenerate in the corresponding eigenspectrum, then $\bra{\phi_k^{(0)}} R \ket{\phi_k^{(0)}} = 0$.
Hence for such initial states, Eq.~(\ref{eq:MRate}) gives the initial rate of change of $M$ as $0$.

In the case of eigenvalue degeneracies, it is natural to consider an initial ensemble where the states $\{ \ket{\phi_k^{(0)}} \}$ appear with probabilities that depend only on their $H^{(0)}$ and $M^{(0)}$ eigenvalues, $\{ p_k = f(\epsilon_k^{(0)}, \mu_k^{(0)}) \}$.
For such an ensemble, using Eq.~(\ref{eq:Rmatrelem}) we can easily see that
\begin{equation}
\sum_k p_k \bra{\phi_k^{(0)}} R \ket{\phi_k^{(0)}} = 0 ~,
\end{equation}
hence the initial rate of change of the observable $M$ is $0$.

Note that in the above we have only used the quasi-commutation condition Eq.~(\ref{eq:HMcommutator}), specialized to Eq.~(\ref{eq:H0M1_M0H1}) in the present case.
Suppose we further know that the quasi-commutation derives from a truncated unitary rotation, i.e., there exists $X$ such that
\begin{equation}
H^{(1)} = i [X, H^{(0)}] ~, \quad 
M^{(1)} = i [X, M^{(0)}] ~.
\end{equation}
A straightforward calculation utilizing the commutation of $H^{(0)}$ and $M^{(0)}$ and the common eigenstate condition on $\ket{\phi_k^{(0)}}$ then gives $\bra{\phi_k^{(0)}} R \ket{\phi_k^{(0)}} = 0$ without the non-degeneracy assumption.
Indeed, we simply write out
\begin{align}
i R &= X H^{(0)} X M^{(0)} - H^{(0)} X^2 M^{(0)} + H^{(0)} X M^{(0)} X \label{eq:Rwriteout} \\
&~~ - X M^{(0)} X H^{(0)} + M^{(0)} X^2 H^{(0)} - M^{(0)} X H^{(0)} X \nonumber 
\end{align}
and observe that the above terms can be grouped in pairs that cancel each other when evaluated on any common eigenstate $\ket{\phi_k^{(0)}}$, e.g., the first and the last term, etc.

A remark is in order.
When $X$ is a boosted or a bi-local operator, while we can use it in an infinite system to generate such a quasi-commuting Hamiltonian and IoMs, it is not clear if we can turn this argument into a proper proof in a finite system.
On the other hand, the earlier arguments with the additional eigenvalue non-degeneracy assumptions, or with appropriately generalized initial ensembles, still work.
We presented the argument with the truncated unitary rotation generated by $X$ since its manipulations are purely algebraic and extend more simply to some higher-order calculations below, and we conjecture that the conclusions still hold even when we cannot define such $X$ in a finite system.

\subsection{Perturbative treatment for the dynamics of $M$}
We can evaluate the R.H.S.\ of Eq.~(\ref{eq:MRate}) perturbatively in $\lambda$ using
\begin{align*}
& e^{iHt} R e^{-iHt} \approx e^{iH^{(0)}t} R e^{-iH^{(0)}t} + \\
& + i \lambda \int_0^t ds e^{i H^{(0)} (t-s)} [H^{(1)}, e^{i H^{(0)} s} R e^{-i H^{(0)} s}] e^{-i H^{(0)} (t-s)}
\end{align*}
valid to formal $O(\lambda^2)$.
When evaluated in an initial state $\ket{\Psi_0}$ that is an eigenstate of $H^{(0)}$, this gives
\begin{align}
& \bra{\Psi_0} e^{iHt} R e^{-iHt} \ket{\Psi_0} \approx \bra{\Psi_0} R \ket{\Psi_0} + \\
& + i \lambda \int_0^t ds \bra{\Psi_0} [H^{(1)}, e^{i H^{(0)} s} R\, e^{-i H^{(0)} s}] \ket{\Psi_0} ~.
\label{eq:Rdynamics}
\end{align}
Assuming further that $\ket{\Psi_0}$ is an eigenstate of $M^{(0)}$, with additional natural assumptions considered in the previous subsection,
we showed that $\bra{\Psi_0} R \ket{\Psi_0} = 0$.
Hence the leading term in the formal powers of $\lambda$ series for the R.H.S.\ of Eq.~(\ref{eq:MRate}) is $O(\lambda^3)$.

\subsubsection{Connection with direct perturbative treatment for the dynamics of $M^{(0)}$ and vanishing ``Fermi's Golden Rule'' rate}
In a direct perturbative treatment, we would calculate the rate of change of the {\it unperturbed} observable $M^{(0)}$, rather than $M$.  A non-vanishing such rate at large $t$ can be loosely referred to as ``Fermi's Golden Rule'' (FGR) in the many-body setting \cite{Mallayya2019, Mallayya2021}.
For reference and comparison with literature, we list expression for the $O(\lambda^2)$ term in $\frac{d}{dt} \bra{\Psi(t)} M^{(0)} \ket{\Psi(t)}$ obtained in the direct perturbation theory simplified using existence of $M^{(1)}$ satisfying Eq.~(\ref{eq:H0M1_M0H1}):
\begin{align}
& \frac{d}{dt} \bra{\Psi(t)} M^{(0)} \ket{\Psi(t)} = \nonumber \\
& = i\lambda^2 \bra{\Psi_0} \![H^{(1)}, M^{(1)} - e^{iH^{(0)}t} M^{(1)} e^{-iH^{(0)}t}]\! \ket{\Psi_0} + O(\lambda^3). \label{eq:M0rate_lambda2}
\end{align}
In our treatment, the first (time-independent) term vanishes by Eq.~(\ref{eq:Rval0}), while we next focus on the second term.
Note that the same expression for the leading order in the formal expansion in powers of $\lambda$ can be obtained from $M^{(0)} = M - \lambda M^{(1)}$ and using that the above-calculated $\frac{d}{dt} \bra{\Psi(t)} M \ket{\Psi(t)}$ is formally $O(\lambda^3)$:
\begin{align*}
& \frac{d}{dt} \bra{\Psi(t)} M^{(0)} \ket{\Psi(t)} = -\lambda \frac{d}{dt} \bra{\Psi(t)} M^{(1)} \ket{\Psi(t)}+O(\lambda^3)\\
&  = -i \lambda^2 \bra{\Psi_0} [H^{(1)}, e^{iHt} M^{(1)} e^{-iHt}] \ket{\Psi_0} + O(\lambda^3)  \\
& \approx -i \lambda^2 \bra{\Psi_0} [H^{(1)}, e^{iH^{(0)}t} M^{(1)} e^{-iH^{(0)}t}] \ket{\Psi_0} + O(\lambda^3). ~
\end{align*}

The time-dependent $O(\lambda^2)$ term can be expressed in terms of dynamical correlation functions in the unperturbed problem:
Writing $H^{(1)} = \sum_j h^{(1)}_j$ and $M^{(1)} = \sum_j m^{(1)}_j$,
we have contributions of the form
\begin{align}
& \bra{\Psi_0} [h^{(1)}_j, e^{iH^{(0)}t} m^{(1)}_{j'} e^{-iH^{(0)}t}] \ket{\Psi_0} \label{eq:h1m1commdyn} \\
& = \bra{\Psi_0} h_j^{(1)} e^{iH^{(0)}t} m_{j'}^{(1)} e^{-iH^{(0)}t} \ket{\Psi_0} \nonumber \\
& ~~~ - \bra{\Psi_0} e^{iH^{(0)}t} m_{j'}^{(1)} e^{-iH^{(0)}t} h_j^{(1)} \ket{\Psi_0} ~, \nonumber
\end{align}
which is a difference of the specific dynamical correlation functions.

At this point, if we assume ``factorization'' of dynamical correlation functions of local observables $a_j$ and $b_{j'}$ at large $t$, 
\begin{align}
& \bra{\Psi_0} a_j e^{iH^{(0)}t} b_{j'} e^{-iH^{(0)}t} \ket{\Psi_0} \label{eq:abdyncorr} \\
& \approx \bra{\Psi_0} a_j \ket{\Psi_0} \, \bra{\Psi_0} e^{iH^{(0)}t} b_{j'} e^{-iH^{(0)}t} \ket{\Psi_0} \nonumber \\
& = \bra{\Psi_0} a_j \ket{\Psi_0} \, \bra{\Psi_0} b_{j'} \ket{\Psi_0} ~, \nonumber
\end{align}
and similarly for $\bra{\Psi_0} e^{iH^{(0)}t} b_{j'} e^{-iH^{(0)}t} a_j \ket{\Psi_0}$,
then the two parts in Eq.~(\ref{eq:h1m1commdyn}) cancel each other at large $t$ and such contributions vanish.
That is, we assume that the corresponding connected correlation functions decay to zero at large $t$, and we also need this decay to be sufficiently fast for the next step (see below for a discussion of these assumptions).
If we further assume that the correlations also decay sufficiently fast in real space, we could conjecture that
\begin{align}
\sum_{j'} \bra{\Psi_0} [a_j, e^{iH^{(0)}t} b_{j'} e^{-iH^{(0)}t}] \ket{\Psi_0} \approx 0 \label{eq:sumabdyncorr} 
\end{align}
at large $t$ and would conclude that the $O(\lambda^2)$ contribution to the rate of change of $m^{(0)}_j$ vanishes at large $t$ [where we went from the extensive local operator $M^{(0)}$ to its local part in the spirit of Eq.~(\ref{eq:evol_localm})].
Note that the calculated quantity is solely a property of the unperturbed Hamiltonian $H^{(0)}$ and the initial state $\ket{\Psi_0}$ (or the initial ensemble $\rho_0$ in general).
As such, ``large enough $t$'' is in the units of (inverse) energy scale of $H^{(0)}$, which we take to be $O(1)$.
So the conjectured conclusion is that the formal $O(\lambda^2)$ rate of change of $M^{(0)}$ vanishes after $O(1)$ time.

\subsubsection{Assumptions about the dynamical correlation functions}
Let us be more specific about the precise conjecture used to argue the vanishing of FGR rate and various assumptions motivating it.

First, because of the locality of $H^{(0)}$ (which we always assume), the Lieb-Robinson bound implies that at any fixed $t$ the dynamical correlation functions like Eq.~(\ref{eq:abdyncorr}) are well defined in the thermodynamic limit, and that with such correlation functions the L.H.S.\ of Eq.~(\ref{eq:sumabdyncorr}) is a convergent sum over the real-space separations $|j'-j|$; this means that we have a well-defined formal $O(\lambda^2)$ FGR rate of change of the local observable $m_j^{(0)}$ at any $t$ in the thermodynamic limit. Our main conjecture is that the sum in Eq.~(\ref{eq:sumabdyncorr}) --- and hence this FGR rate --- vanishes for sufficiently large $t$.

Next, we discuss the underlying assumptions about the dynamical correlation functions.
While the ``factorization'' in Eq.~(\ref{eq:abdyncorr}) (i.e., vanishing of the connected correlation functions) is intuitively reasonable, we do not know of a proof in general.
Several recent papers~\cite{Alhambra2020, Watanabe2020, Huang2019} obtained some rigorous results about such ``factorization'' for general Hamiltonians.
However, these results are either in different regimes (e.g., correlations of local observables in the $t \to \infty$ limit first before the thermodynamic limit), or are too weak for us to use (e.g., bounds on  correlations of extensive local observables).
Nevertheless, our situation is better in that we actually need the specific difference of correlation functions which corresponds to the commutator in Eq.~(\ref{eq:abdyncorr}).
For fixed $t$ and large enough $|j'-j| > v_{\text{LR}} t$, i.e., outside the Lieb-Robinson cone where $v_{\text{LR}}$ is the corresponding velocity, we expect this to decay exponentially with $|j'-j|$, so the sum in Eq.~(\ref{eq:sumabdyncorr}) is convergent at any $t$.
Whether this sum indeed approaches zero at large $t$ depends on the behavior of the correlation functions within the Lieb-Robinson cone $|j'-j| \lesssim v_{\text{LR}} t$.
Unfortunately, we do not know sufficiently strong general results for such correlations.
Still, if we can establish that, e.g., correlations at the same location $j'=j$ decay sufficiently quickly with time, it is reasonable to assume that the total contribution from all $|j'-j| \lesssim v_{\text{LR}} t$ also decays to zero with time.

Furthermore, as already mentioned, these are solely questions about the unperturbed $H^{(0)}$ and may be directly answerable for solvable $H^{(0)}$ like all the integrable models considered in this paper.
Thus, for free-fermion unperturbed Hamiltonians like the ones in Sec.~\ref{sec:ex_biloc}, either for the ground state or thermal initial ensembles, one can employ Wick's theorem and carry out such calculations exactly.
While we have not done the full calculations for the corresponding near-integrable perturbations $H^{(1)}$ and $M^{(1)}$ as observables, toy calculations with simpler local observables (e.g., $a_j = c_j^\dagger c_{j+1}^\dagger$, $b_{j'} = c_{j'} c_{j'+1}$, for the spinless fermion hopping problem) suggest that the above conjecture indeed holds, and we leave full calculations for future work.

For interacting integrable models, Ref.~\cite{Jung2006} studied thermal conductivity in the nearest-neighbor Heisenberg chain perturbed by the second-neighbor Heisenberg terms; using numerical high-temperature series and exact diagonalization estimates they found that the corresponding FGR-like $O(\lambda^2)$ contribution indeed vanishes in this case.
Also, Ref.~\cite{Durnin2021} appealed to the hydrodynamic projection principle calculations of the dynamical correlation functions in integrable models to argue that the FGR-like rate of change of an unperturbed IoM indeed vanishes after some initial time, for special perturbations that are equivalent to the ones in our work obtained using generators $X$ that are boosted operators.
More precisely, one can start with Eq.~(\ref{eq:M0rate_lambda2}) and use relation $\bra{\Psi_0}[H^{(1)}, M^{(1)} - e^{iH^{(0)}t} M^{(1)} e^{-iH^{(0)}t}] \ket{\Psi_0} = \bra{\Psi_0} H^{(1)} (e^{-iH^{(0)}t} M^{(1)} e^{iH^{(0)}t} - e^{iH^{(0)}t} M^{(1)} e^{-iH^{(0)}t}) \ket{\Psi_0}$, which can be checked, e.g., by differentiating both sides with respect to $t$ and utilizing Eq.~(\ref{eq:H0M1_M0H1}).
The latter form matches expression analyzed in Ref.~\cite{Durnin2021}, which then appealed to the hydrodynamic projection principle to suggest that the corresponding positive time and negative time   correlations effectively cancel each other at large $t$.

While these calculations are specific for integrable models, we think that our conjecture does not require integrability and holds for generic local Hamiltonians $H^{(0)}$, perhaps under mild additional conditions, but leave more tests (e.g., by numerical methods) to future work.

Finally, we emphasize that while fully justifying the vanishing of the $O(\lambda^2)$ rate for the $M^{(0)}$ required additional arguments like the ones in this subsection, no such arguments were needed when considering the $O(\lambda^2)$ rate for the properly corrected (i.e., quasi-conserved) observable $M$.

\subsubsection{Vanishing of the $O(\lambda^3)$ term in the rate of change of $M$ at long time}
Returning to the dynamics of the full $M$ and Eq.~(\ref{eq:Rdynamics}), in the case $H^{(1)} = i [X, H^{(0)}]$ and using that $\braket{\Psi_0|[i[X,H^{(0)}], \bullet]|\Psi_0} = \braket{\Psi_0|[X,i[H^{(0)}, \bullet]]|\Psi_0}$, we have 
\begin{align}
& \bra{\Psi_0} [H^{(1)}, e^{i H^{(0)} s} R e^{-i H^{(0)} s}] \ket{\Psi_0} \\
& = \bra{\Psi_0} [X, i [H^{(0)}, e^{i H^{(0)} s} R e^{-i H^{(0)} s}]] \ket{\Psi_0} \nonumber \\
& = \bra{\Psi_0} [X, \frac{d}{ds} \left( e^{i H^{(0)} s} R e^{-i H^{(0)} s} \right) ] \ket{\Psi_0} ~, \nonumber
\end{align}
and hence
\begin{align}
\frac{d}{dt}& \bra{\Psi(t)} M \ket{\Psi(t)} =\nonumber\\
&=i \lambda^3 \bra{\Psi_0} [X, e^{i H^{(0)} t} R e^{-i H^{(0)} t} - R] \ket{\Psi_0} + O(\lambda^4) ~.
\end{align}
Using Eq.~(\ref{eq:Rwriteout}) and writing out $[X, R]$, we can similarly pair terms such that their expectation values in the state $\ket{\Psi_0}$ cancel each other, giving
\begin{equation}
\bra{\Psi_0} [X, R] \ket{\Psi_0} = 0 ~.
\label{eq:XRval0}
\end{equation}
Furthermore, if we assume ``factorization'' of the dynamical correlation functions at long times, similar to Eq.~(\ref{eq:abdyncorr}) and conjecture Eq.~(\ref{eq:sumabdyncorr}), we then have
\begin{equation}
\bra{\Psi_0} [X, e^{i H^{(0)} t} R e^{-i H^{(0)} t}] \ket{\Psi_0} \approx 0
\end{equation}
for large $t$.
Hence after some initial time, the formal $O(\lambda^3)$ term in the rate of change of $M$ vanishes.
The non-vanishing rate is then $O(\lambda^4)$, in agreement with the expectations of FRG reasoning applied to an effective strength of true integrability breaking being $O(\lambda^2)$.

Some remarks are in order.
Note that we also needed such arguments appealing to ``factorization'' of the dynamical correlations to show the vanishing of the $O(\lambda^2)$ rate of change of $M^{(0)}$ after some initial time, while we did not need such arguments for the $O(\lambda^2)$ rate of change of the ``corrected'' (quasi-IoM) $M$---the formal $O(\lambda^2)$ rate of change of $M$ is identically zero at any time.
We think that we need such arguments for the $O(\lambda^3)$ rate of change of the corrected $M$ because the initial state $\ket{\Psi_0}$ is not ``corrected'' to reflect that the true integrability breaking perturbation has effective strength $O(\lambda^2)$.

Indeed, let us return to Eq.~(\ref{eq:MRate}) valid for any initial state, and instead of $\ket{\Psi_0}$ [a common eigenstate of $H^{(0)}$ and $M^{(0)}$ as before], let us start with
\begin{equation}
\ket{\Psi_{\text{ini}}} = e^{i \lambda X} \ket{\Psi_0} \quad \text{or} \quad
\rho_{\text{ini}} = e^{i\lambda X} \rho_0 e^{-i\lambda X}
\end{equation}
(assuming as before that $[\rho_0, H^{(0)}] = [\rho_0, M^{(0)}] = 0$).
In this case we have
\begin{equation}
\frac{d}{dt} \bra{\Psi(t)} M \ket{\Psi(t)} = \lambda^2 \bra{\Psi_0}
e^{i \tilde{H} t}
\tilde{R} e^{-i \tilde{H} t} \ket{\Psi_0}
\end{equation}
with
\begin{align}
& \tilde{H} \equiv e^{-i\lambda X} H e^{i\lambda X} = H^{(0)} + O(\lambda^2) ~, \\
& \tilde{R} \equiv e^{-i\lambda X} R e^{i\lambda X} = R - i \lambda [X,R] + O(\lambda^2) ~.
\end{align}
Hence
\begin{align}
& \frac{d}{dt}\! \bra{\Psi(t)} \!M\! \ket{\Psi(t)} \!=\! \lambda^2 \bra{\Psi_0}\! e^{i H^{(0)} t} \tilde{R} e^{-i H^{(0)} t} \!\ket{\Psi_0} + O(\lambda^4) \nonumber \\
& = \lambda^2 \bra{\Psi_0}
(R - i \lambda[X,R]) \ket{\Psi_0} + O(\lambda^4) = O(\lambda^4),
\end{align} 
where the $O(\lambda^2)$ and $O(\lambda^3)$ terms vanish by Eqs.~(\ref{eq:Rval0}) and (\ref{eq:XRval0}).
Thus, for such initial states or ensembles, the leading contribution to the rate of change of $M$ is $O(\lambda^4)$ at any time $t$.

One may worry that such $\ket{\Psi_{\text{ini}}}$ or $\rho_{\text{ini}}$ are not readily preparable (e.g., experimentally preparable) from $\ket{\Psi_0}$ or $\rho_0$.
However, if $X$ is a sum of on-site terms (as in the example in Sec.~\ref{sec:conclusions} of the on-site Hubbard term used as a generator starting with free fermions in any dimension), then $e^{i \lambda X}$ is a product of on-site unitaries and therefore is a depth-1 unitary circuit, hence such initial states or ensembles can be viewed as preparable from the unperturbed ones.
For a general extensive local $X$, we can appropriately Trotterize $e^{i \lambda X}$ and obtain a finite depth unitary circuit $U'(\lambda)$ that reproduces $e^{i \lambda X}$ with $O(\lambda^2)$ accuracy (e.g., $e^{i \lambda X} \approx e^{i\lambda X_{\text{even bonds}}} e^{i\lambda X_{\text{odd bonds}}} \equiv U'(\lambda)$ familiar for a 1d chain with only nearest-neighbor bond terms in $X$); in this case we can initialize with the corresponding $\ket{\Psi_{\text{ini}}'} = U'(\lambda) \ket{\Psi_0}$ or $\rho_{\text{ini}}' = U'(\lambda) \rho_0 U'(\lambda)^\dagger$ differing from the above $\ket{\Psi_{\text{ini}}}$ or  $\rho_{\text{ini}}$ by $O(\lambda^2)$, and observe that with such preparable initial states or ensembles we also obtain vanishing formal $O(\lambda^2)$ and $O(\lambda^3)$ terms in the rate of change of $M$.
[Note that these $\ket{\Psi_{\text{ini}}'}$ or $\rho_{\text{ini}}'$ are \emph{not} the naive truncated series approximations $\ket{\Psi_0} + i \lambda X \ket{\Psi_0}$ or $\rho_0 + i \lambda [X, \rho_0]$ that would be problematic since $X$ is an extensive operator.]

One should worry more about the case where the generator $X$ is a boosted or bilocal operator, hence is not a regular local extensive operator.
While we do not have a full resolution of this concern, we observe that the commutator of such $X$ with $H^{(0)}$ and $M^{(0)}$ gives regular extensive local operators; hence, the non-locality of $X$ is mitigated when one considers, e.g., $\rho_0$ that is an equilibrium ensemble under $H^{(0)}$ and $M^{(0)}$:
\begin{align}
& \rho_0 \equiv \frac{1}{Z_0} e^{-\beta H^{(0)} - \gamma M^{(0)}} ~, \quad Z_0 \equiv \text{Tr}\Big(e^{-\beta H^{(0)} - \gamma M^{(0)}}\Big) ~, \nonumber \\
& \implies \rho_{\text{ini}} = \frac{1}{Z_0} e^{-\beta e^{i\lambda X} H^{(0)} e^{-i\lambda X} - \gamma e^{i\lambda X} M^{(0)} e^{-i\lambda X}} \nonumber \\
& \approx \frac{1}{Z_0} e^{-\beta (H^{(0)} + \lambda H^{(1)}) - \gamma (M^{(0)} + \lambda M^{(1)})} + O(\lambda^2) ~,
\end{align}
where $O(\lambda^2)$ means in formal series in $\lambda$.
In particular, this implies that in such series
\begin{align}
& Z_0' \equiv \text{Tr}\Big( e^{-\beta (H^{(0)} + \lambda H^{(1)}) - \gamma (M^{(0)} + \lambda M^{(1)})} \Big)\approx Z_0 + O(\lambda^2) ~, \nonumber \\
& \rho_{\text{ini}}' \equiv \frac{1}{Z_0'} e^{-\beta (H^{(0)} + \lambda H^{(1)}) - \gamma (M^{(0)} + \lambda M^{(1)})} = \rho_{\text{ini}} + O(\lambda^2) ~.
\end{align}
Hence, for such $\rho_{\text{ini}}'$ we expect the formal rate of change of $M$ is $O(\lambda^4)$.
Note that this is \emph{not} an equilibrium ensemble for $H = H^{(0)} + H^{(1)}$, since $H$ does not commute exactly with $M = M^{(0)} + M^{(1)}$.
Intuitively, we expect such ensembles to describe prethermalized states in our system with quasi-conserved $M$, while here we use $\rho_{\text{ini}}'$ to illustrate initial ensembles that will ``know'' from the outset that the true integrability breaking strength is $O(\lambda^2)$ and where any reference to $X$ has dropped out.

Indeed, we can verify this by a direct calculation:
\begin{align}
& \frac{d}{dt} \text{Tr}\left(\rho_\text{ini}' e^{iHt} M e^{-iHt} \right) = \lambda^2 \text{Tr} \left(\rho_\text{ini}' e^{iHt} R e^{-iHt} \right) \\
& = \lambda^2 \text{Tr}\left(\rho_\text{ini}' R\right) + O(\lambda^4) ~, 
\end{align}
since $H$ and the operator in the exponent of $\rho_\text{ini}'$ commute to $O(\lambda^2)$.
Furthermore, using $[\rho_\text{ini}', \beta H + \gamma M] = 0$, we have
\begin{align}
& \text{Tr}\left(\rho_\text{ini}' \lambda^2 R\right) = \text{Tr}\left(\rho_\text{ini}' i[H, M] \right) 
= i\text{Tr}\big(\rho_\text{ini}' \big[H + \frac{\gamma}{\beta}M, M\big] \big) \nonumber \\
& = i\text{Tr}\big(\big[\rho_\text{ini}', H + \frac{\gamma}{\beta}M \big] M \big) = 0 ~.
\end{align}
Hence, for such initial ensembles, the rate of change of $M$ at any time $t$ is formally $O(\lambda^4)$.

It is not clear if this $\rho_{\text{ini}}'$ can be viewed as preparable from $\rho_0$ by a finite depth local unitary circuit and perhaps a better treatment can be found.
Here were are content with the above arguments how to more readily see that the leading ``thermalization'' rate in perturbation theory, after some initial relaxation, should indeed be set by $O(\lambda^4)$.

The above readily generalizes to a situation with multiple IoMs that become quasi-IoMs upon adding special perturbations.
In this case in the uperturbed problem we can consider $\rho_0$ in the form of a Generalized Gibbs Ensemble
\begin{equation}
\rho_0 = \rho_\text{GGE}\propto \exp\left(-\sum_\alpha \gamma_\alpha Q_\alpha^{(0)}\right) ~,
\end{equation}
where $H^{(0)}$ is included as one of the $Q_\alpha$'s.
A suitable initial state with formal rate of change of order $O(\lambda^4)$ in all the quasi-IoMs can then be obtained as
\begin{align}
\rho_\text{ini}' & \propto \exp\left[-\sum_\alpha \gamma_\alpha (Q_\alpha^{(0)}+\lambda Q_\alpha^{(1)}) \right] ~.
\end{align}

\end{document}